\documentclass[preprints,article,accept,moreauthors,pdftex]{Definitions/mdpi} 
\firstpage{1} 
\pubvolume{9}
\issuenum{7}
\articlenumber{996}
\pubyear{2022}
\copyrightyear{2022}
\externaleditor{Academic Editor: Jareen K. Meinzen-Derr
}
\datereceived{26 May 2022} 
\dateaccepted{30 June 2022} 
\datepublished{1 July 2022} 
\hreflink{https://doi.org/10.3390/ \linebreak children9070996} 

\usepackage{CJKutf8}
\usepackage{lscape}
\usepackage{xltabular}
\usepackage[tone]{tipa}
\usepackage[labelformat=simple]{subcaption}
\DeclareCaptionLabelFormat{subcaptionlabel}{\normalfont(\textbf{#2}\normalfont)}
\graphicspath{{media/}}

\Title{Deep-Learning-Based Automated Classification of Chinese Speech Sound Disorders}

\TitleCitation{Deep-Learning-Based Automated Classification of Chinese Speech Sound Disorders}

\Author{Yao-Ming Kuo $^{1}$\orcidA{}, Shanq-Jang Ruan $^{1}$\orcidB{}, Yu-Chin Chen $^{2}$ and Ya-Wen Tu $^{2,}$*}

\AuthorNames{Yao-Ming Kuo, Shanq-Jang Ruan, Yu-Chin Chen and Ya-Wen Tu}

\AuthorCitation{Kuo, Y.-M.; Ruan, S.-J.; Chen, Y.-C.; Tu, Y.-W.}

\address{%
$^{1}$ \quad Department of Electronic and Computer Engineering, National Taiwan University of Science and Technology, Taipei 106, Taiwan; m10902405@mail.ntust.edu.tw (Y.-M.K.); sjruan@mail.ntust.edu.tw (S.-J.R.)\\
$^{2}$ \quad Sijhih Cathay General Hospital, New Taipei 221, Taiwan; cgh398506@cgh.org.tw}

\corres{Correspondence: yawentu@gmail.com}

\abstract{This article describes a system for analyzing acoustic data  to assist in the diagnosis and classification of children's speech sound disorders (SSDs) using a computer. The analysis concentrated on identifying and categorizing four distinct types of Chinese SSDs. The study collected and generated a speech corpus containing 2540 stopping, backing, final consonant deletion process (FCDP), and affrication samples from 90 children aged 3--6 years with normal or pathological articulatory features. Each recording was accompanied by a detailed diagnostic annotation by two speech--language pathologists (SLPs). Classification of the speech samples was accomplished using three well-established neural network models for image classification. The feature maps were created using three sets of Mel-frequency cepstral coefficients (MFCC) parameters extracted from speech sounds and aggregated into a three-dimensional data structure as model input. We employed six techniques for data augmentation to augment the available dataset while avoiding overfitting. The experiments examine the usability of four different categories of Chinese phrases and characters. Experiments with different data subsets demonstrate the system's ability to accurately detect the analyzed pronunciation disorders. The best multi-class classification using a single Chinese phrase achieves an accuracy of 74.4~percent.}


\begin{document}
\section{Introduction\label{1}}

Speech sound disorders (SSDs) are one of the most common disorders in preschool and school-age children. Any issue or combination of difficulties with perception, motor production, or~phonological representation of speech sounds and speech segments—including phonotactic rules controlling allowable speech sound sequences in a language is referred to as an SSD. According to a 2012 National Center for Health Statistics study~\cite{black2015communication}, 48.1~percent of 3- to 10-year-old children and 24.4~percent of 11- to 17-year-old children with a communication impairment had just speech sound difficulties. Children with speech difficulties had a 76.6~percent use rate of speech intervention services, as~reported by their parents~\cite{black2015communication}. Based on~\cite{wren2016prevalence}, speech delay or SSDs affect 2.3~percent to 24.6~percent of school-aged~children.

\textls[-15]{There are two types of SSDs: organic and functional. An underlying motor/neurological, structural, or~sensory/perceptual reason causes organic SSDs. There is no known cause for functional speech sound disorders; they are idiopathic. Functional SSDs are divided into two categories: motor production of speech and linguistic aspects of speech production. These issues have been referred to as articulation and phonological disorders, respectively, in~the past. Errors (such as distortions and replacements) in producing particular speech sounds focus on articulation disorders. Phonological disorders are characterized by predictable, rule-based mistakes that influence several sounds (e.g., fronting, backing, and~final consonant deletion).\cite{ASHA}}

When a child has poor intelligibility, parents can visit a rehabilitation clinic and then be referred to SLPs for examination and training following assessment. According to~\cite{Chang2019Assessment}, it takes an average of 54 min per case for assessment and analysis. Because~there is a shortage of speech--language pathologists (SLPs) in Taiwan~\cite{Sen2017Study}, children with SSDs often have to spend a longer waiting time visiting a clinic or the rehabilitation department of a medical institution. The~waiting period is also a golden opportunity to miss out on treatment. Moreover, the~lack of clarity in children's speech can easily affect children's social and communication interactions. Some children's poor mastery of phonological rules can affect their future phonetic or intonation awareness~\cite{rvachew2007phonological}. According to the literature, speech therapy effectively improves children's condition if started early~\cite{eadie2015speech}. The~diagnosis of speech sounds varies depending on the method or location of the speech sound, so we can classify and model the features of the speech sound into specific categories. Correct diagnosis of pronunciations is the first step in clinical treatment, as~the elicitation techniques vary by class. However, now in Taiwan, there is a lack of standardized assessment tools. The~evaluation procedure may differ from one SLP to another due to differences in auditory awareness and not having  a standard evaluation tool. Furthermore, as~there are no normative models to compare evaluated instances to, it is difficult to make meaningful comparisons between them. Additionally, the~assessment content varies from monotone vocabulary to spontaneous speech. The~overall workflow is lengthy and laborious, and~therapists are frequently required to complete the assessment and health education in less than 30 min, which is exhausting and inconvenient. Therefore, the~availability of automatic classification assessment tools can save time for SLPs and quickly identify speech problems in children and provide accurate treatment~directions.

\subsection{Disorders~Characterizations}
The phonological processes are divided into syllabic structure, substitution, and~assimilation. Substitution processes can be classified by their articulation method or location. The~term “place of articulation” refers to the point at which two speech organs, such as the tongue and teeth, come into contact to produce speech sounds. The~manner in which the articulatory structures are shaped and coordinated determines the manner in which they articulate, and~common diagnoses such as stopping and affrication are extremely diverse. To~create different speech sounds, we experimented with various airflow methods, the~degree of airflow obstruction, and~the duration of airflow. According to~\cite{jeng2011phonological}, the~most common types of errors in preschool children are backing, stopping, affrication, and~unaspiration. The~current study focuses on four types of errors that are frequently encountered: stopping, backing, final consonant deletion process (FCDP), and~affrication.

Using spectrograms to analyze speech problems can reveal a wealth of information that cannot be analyzed by the ear. The~horizontal axis of the spectrogram is the time scale, and the vertical axis is the frequency of the sound. The~vertical axis is the frequency of the sound, and~from the bottom to the top is the logarithmic scale from 0 to 20,000 Hz, which represents the range of audible sound. Using a logarithmic scale emphasizes the range of frequencies emitted by the vocal cords. The~spectrum's brightness indicates the sound's magnitude at the corresponding time and frequency. The~higher the dB value, the~brighter the color, and~the lower the dB value, the~darker the~color.

\begin{CJK*}{UTF8}{bsmi}
	\subsubsection{Stopping}
\begin{figure}[H]
		{\captionsetup{position=bottom,justification=centering}\begin{subfigure}{.29\textwidth}
			\includegraphics[height=.95\linewidth]{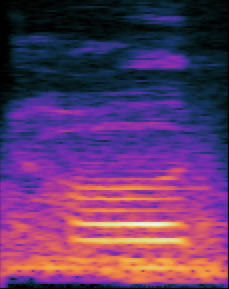}
			\caption{}
			\label{fig:subs1}
			\end{subfigure}\hspace{-0.3cm}
			\begin{subfigure}{.29\textwidth}
				\includegraphics[height=.95\linewidth]{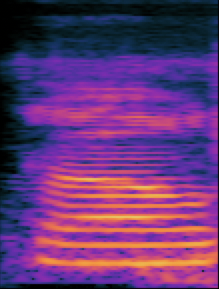}
				\caption{}
				\label{fig:subs2}
			\end{subfigure}}
		\caption{Spectrum comparison of stopping and~correct: (\textbf{a}) stopping; (\textbf{b}) correct.}
		\label{fig:tests}
	\end{figure}

	Stopping refers to when non-stop sounds are incorrectly pronounced as stop; in Chinese, stop sounds include ㄅ/p/, ㄆ\textipa{/p\textsuperscript{h}/}, ㄉ/t/, ㄊ\textipa{/t\textsuperscript{h}/}, ㄍ/k/, and~ㄎ\textipa{/k\textsuperscript{h}/}; therefore, when we mispronounce other sounds into the above six sounds in our daily lives, we will experience stopping. It is referred to as stopping, as~in \textipa{/k\textsuperscript{h}u\tone{55}\tone{11} tsW\tone{33}/} read as \textipa{/tu\tone{55}\tone{11} tsW\tone{33}/}, but~the stop sound contains the two sounds ㄍ/k/ and ㄎ\textipa{/k\textsuperscript{h}/}. When we mispronounce the pronunciation as ㄍ/k/ or ㄎ\textipa{/k\textsuperscript{h}/} speech in clinical practice, we do not refer to it as stopping but rather as backing, as~explained in the following subsection. When the sound spectrum is analyzed, we can see that the stop exhibits the following characteristics. The~first is the duration of silence, which is the duration of the stop being blocked; The time interval between the burst and the beginning of the vowel is referred to as the voice onset time (VOT). We can distinguish various speech sounds based on the acoustic characteristics listed above. Figure~\ref{fig:tests} depicts the spectrogram difference between the stopping and normal~pronunciation.
	
	\subsubsection{Backing}
\begin{figure}[H]
		{\captionsetup{position=bottom,justification=centering}\begin{subfigure}{.29\textwidth}
			\includegraphics[height=0.95\linewidth]{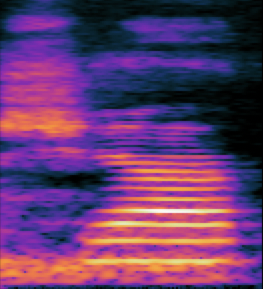}
			\caption{}
			\label{fig:subb1}
			\end{subfigure}%
			\begin{subfigure}{.29\textwidth}
				\includegraphics[height=0.95\linewidth]{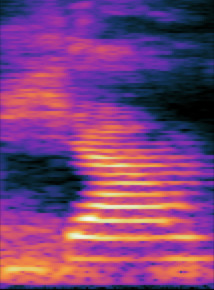}
				\caption{}
				\label{fig:subb2}
			\end{subfigure}}
		\caption{Spectrum comparison of backing and~correct: (\textbf{a}) backing; (\textbf{b}) correct.}
		\label{fig:testb}
	\end{figure}

	The Chinese backing consonants include ㄍ/k/, ㄎ\textipa{/k\textsuperscript{h}/} and ㄏ/x/.When we pronounce Chinese pronunciation, the~stop, affrication, fricative, etc. are replaced by the ㄍ/k/ and ㄎ\textipa{/k\textsuperscript{h}/}, and we call it backing. For~example, \textipa{/t\textsuperscript{h}u\tone{55}\tone{11} tsW\tone{33}/} becomes \textipa{/k\textsuperscript{h}u\tone{55}\tone{11} tsW\tone{33}/}. In~English, backing can occur at any point in the word, but~in Chinese, the~phonological progression of backing occurs exclusively in consonants, and~thus the error occurs at the beginning of the word, which is referred to as the initial consonant in Chinese. The~term “backing” refers to a speech sound produced by the soft palate being held upward by the tongue bulging at the back of the mouth. As~a result, the~acoustic characteristics of stopping are also present in backing, such as silence gap, burst, VOT, and~noise. Figure~\ref{fig:testb} depicts the spectrogram difference between the backing and normal~pronunciation.
	
	\subsubsection{FCDP}
\begin{figure}[H]
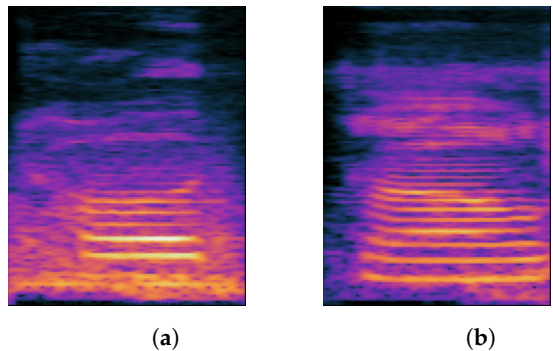

		{\captionsetup{position=bottom,justification=centering}\begin{subfigure}{.29\textwidth}
			\includegraphics[height=.95\linewidth]{16_1.png}
			\caption{}
			\label{fig:subf1}
			\end{subfigure}%
			\begin{subfigure}{.29\textwidth}
				\includegraphics[height=.95\linewidth]{16_16_1.png}
				\caption{}
				\label{fig:subf2}
			\end{subfigure}}
		\caption{Spectrum comparison of FCDP and~correct: (\textbf{a}) FCDP; (\textbf{b}) correct.}
		\label{fig:testf}
	\end{figure}

	The final consonant is composed of a vowel and a coda and~is pronounced by progressing from vowel to consonant. The~final consonant is divided into two segments: the stop coda and the nasal coda. However, only the nasal coda contains the following consonants in the Chinese phonetic alphabet: ㄢ/an/, ㄣ\textipa{/@n/}, ㄤ\textipa{/AN/}, ㄥ\textipa{/\textramshorns N/}. Therefore, the~final consonant is considered as syllable structure component, and~the deletion of the final consonant is referred to as the FCDP. The~following section discusses the final consonant's composition. The~final consonant is categorized by the vowel ㄚ/ä/ or ㄜ\textipa{/\textramshorns/}, followed by /n/ or \textipa{/N/} at the end of the rhyme (coda), which can be roughly divided into two groups: ㄤ\textipa{/AN/}, ㄢ/an/ and ㄣ\textipa{/@n/}, ㄥ\textipa{/\textramshorns N/}. When we pronounce ㄢ/an/, we place our tongue at its lowest point and slowly raise the tip of the tongue, allowing air to flow out of the nasal cavity; when we pronounce ㄤ\textipa{/AN/}, we also place our tongue at its lowest point and slightly open our mouth, allowing air to flow out of the nasal cavity while we keep our mouth open and pronounce the velar nasal \textipa{/N/}. When pronouncing ㄣ\textipa{/@n/} or ㄥ\textipa{/\textramshorns N/}, the~tongue is positioned in the mouth without moving up, down, forward, or~backward, forming the vowel position of ㄜ\textipa{/\textramshorns/}, and~the tongue tip moves up and out through the nasal cavity, producing an alveolar nasal /n/. To~produce a response, on~the other hand, a~vowel position of ㄜ\textipa{/\textramshorns/} is formed first; then the mouth remains open, and the airflows out of the nasal cavity, maintaining the mouth open and producing the velar nasal \textipa{/N/}. Figure~\ref{fig:testf} depicts the spectrogram difference between the FCDP and normal~pronunciation.
	
	\unskip
	
	\subsubsection{Affrication}
\begin{figure}[H]
		{\captionsetup{position=bottom,justification=centering}\begin{subfigure}{.29\textwidth}
			\includegraphics[height=.95\linewidth]{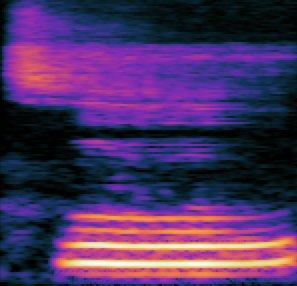}
			\caption{}
			\label{fig:suba1}
			\end{subfigure}\,\,\,%
			\begin{subfigure}{.29\textwidth}
				\includegraphics[height=.95\linewidth]{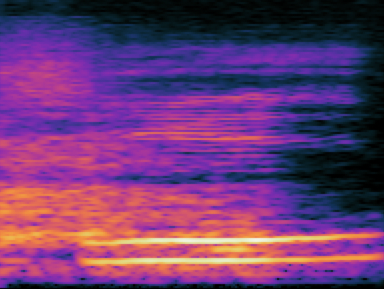}
				\caption{}
				\label{fig:suba2}
			\end{subfigure}}
		\caption{Spectrum comparison of affrication and~correct: (\textbf{a}) affrication; (\textbf{b}) correct.}
		\label{fig:testa}
	\end{figure}

	An affricate contains both stop and fricative features, so when it is pronounced, the~oral constellation will first produce the stop feature and then the fricative feature. In~Chinese pronunciation, there are six affricates: ㄗ\textipa{/\texttoptiebar{ts}/}, ㄘ\textipa{/\texttoptiebar{\:t\:s}\textsuperscript{h}/}, ㄓ\textipa{/\texttoptiebar{\:t\:s}/}, ㄔ\textipa{/\texttoptiebar{ts}\textsuperscript{h}/}, ㄐ\textipa{/\texttoptiebar{tC}/} and ㄑ\textipa{/\texttoptiebar{tC}\textsuperscript{h}/}. When other phonemes are mispronounced as the six phonemes listed above, they become affrication. The~so-called affricate is a closed tone that lasts for a period of time, forming a block and holding it. However,~during the burst, the~mouth does not completely release the airflow, or~rather forms a small gap between the tongue and the hard palate, allowing the airflow to pass through the gap and produce a friction noise. When we examine the spectrogram, we can see that the affricate consonant has the acoustic characteristics of both the stop consonant and the fricative consonant, such as a silent period, a~burst, and~a short noise. However, the~characteristics of the stop consonant are very dynamic, as~they can change quickly and dramatically, and~we can usually distinguish between them based on this characteristic. Figure~\ref{fig:testa} depicts the spectrogram difference between the affrication and normal~pronunciation.
	
\end{CJK*}

\subsection{State of the~Art}

Despite the enormous potential demand for automatic SSDs classification, some scholars have also researched SSD in different languages. Anjos~et~al.~\cite{anjos2019sibilant} proposed identifying sibilant phonemes in European Portuguese using deep convolutional neural networks. According to~\cite{krecichwost2021automated}, it identified six dental errors in Polish-speaking youngsters using a deep network. Hammami~et~al.~\cite{hammami2020recognition} presented a method based on a real-world database of native Arabic-speaking children's voice recordings. Based on the aforementioned research, it is evident that SSD classification using deep learning is feasible, although~it is currently only used to identify and classify specific single consonants. On~the other hand, relatively few studies have been conducted on Standard Chinese. There are two issues that make detecting the features of different construal errors challenging. First, when growing children attempt to pronounce constantly, the~instability of co-constructive motions manifests. Second, the~numerous features included in a single construal category are diverse, resulting in the difficulty of classification. Recent studies have classified and identified phonetic categories using deep learning architectures. Numerous model architectures are used, such as recurrent neural networks~\cite{wang2018semi}, convolutional neural networks~\cite{lou2018disfluency}, long short-term memory~\cite{wang2017transition}, and~other deep learning frameworks. The~model is fed a two-dimensional spectrogram or Mel-frequency cepstral coefficients (MFCC) data.

\subsection{Aims and~Scope}

Our study aims to develop a reliable data analysis procedure for the computer-assisted diagnosis of SSDs in children. The~goal is to provide a solution of detecting and classifying four types of speech sound errors in Mandarin Chinese. We collected a corpus of speech samples from 90 children aged 3 to 6, along with detailed diagnostic instructions provided by an SLP. The~study is divided into three groups of experiments on pronunciation disorders. We train and compare our gathered dataset for speech sample categorization using three standard architectures: EfficientNet~\cite{tan2019efficientnet}, DenseNet~\cite{iandola2014densenet}, and~InceptionV3~\cite{xia2017inception}. We extract acoustic characteristics from sounds using a three-channel Mel-Spectrogram~\cite{palanisamy2020rethinking}. To~aid the model's learning when trained on custom datasets, we employ various data augmentation techniques~\cite{nanni2020data} on our~dataset.

\subsection{Paper~Structure}

The following is the overall structure of this paper. Section~\ref{2} discusses the methods for pre-use treatment and model training. Section~\ref{3} provides a thorough description of the experimental findings. Section~\ref{4}  discusses the potential reasons why different sound samples influence the accuracy and the bottleneck in the results. Section~\ref{5} concludes the work and future~directions.

\section{Materials and~Methods\label{2}}

\begin{figure}[H]
	\includegraphics[width=12cm]{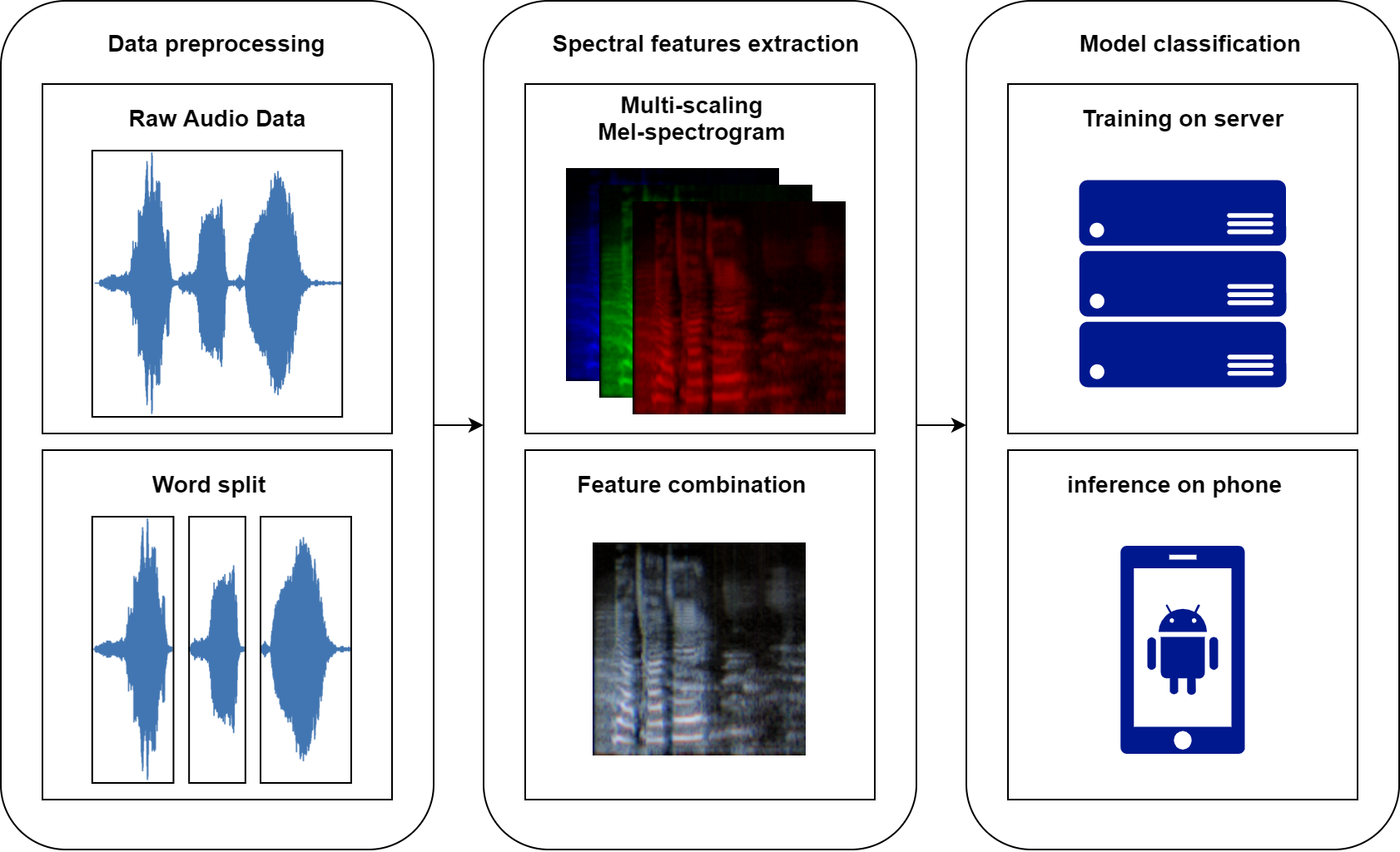}
	\caption{Overall workflow of the speech sound disorders~classification.\label{fig1}}
\end{figure}

The SSD classification task was carried out in accordance with the workflow depicted in Figure~\ref{fig1} and detailed in the following~sections.

\unskip  

\subsection{Collecting and Labeling Audio~Samples}

The study enrolled preschool children aged 3--6 years who had been diagnosed with speech and language impairment at a rehabilitation clinic or were referred by their kindergarten teachers as having a possible speech and language impairment. Between~January and December 2021, a~total of 90 children were enrolled, with~the age and gender distributions shown in Table~\ref{dataagesex}. We excluded cases with the following conditions: speech disorders caused by congenital central nerve injury (e.g., cerebral palsy); speech disorders caused by abnormal oral structures (e.g., cleft lip and palate); co-occurring intellectual disabilities; emotional behavior disorders (e.g., autism); speech disorders caused by hearing impairment; and family reluctance. Prior to the trial, the~protocol was approved by the Cathay Hospital IRB. Consent was obtained verbally and in writing from the child's parents or legal guardians to participate in the~study.

\begin{table}[H]
	\caption{Distribution of subjects' ages and~genders.}
	\label{dataagesex}
	\newcolumntype{C}{>{\centering\arraybackslash}X}
	\begin{tabularx}{\textwidth}{CCCC}
		\toprule
		\multirow{2}{*}{\textbf{Age}}      & \multicolumn{2}{c}{\textbf{Sex}} & \multirow{2}{*}{\textbf{Total}}                          \\
		                          & \textbf{Female}        & \textbf{Male}          &                        \\
		\midrule
		3                         & 8                      & 14                     & 22                     \\
		4                         & 11                     & 18                     & 29                     \\
		5                         & 11                     & 20                     & 31                     \\
		6                         & 4                      & 4                      & 8                      \\
		\midrule
		\multicolumn{1}{c}{Total} & \multicolumn{1}{c}{34} & \multicolumn{1}{c}{56} & \multicolumn{1}{c}{90} \\
		\bottomrule
	\end{tabularx}
\end{table}

\begin{figure}[H]
	\includegraphics[width=8 cm]{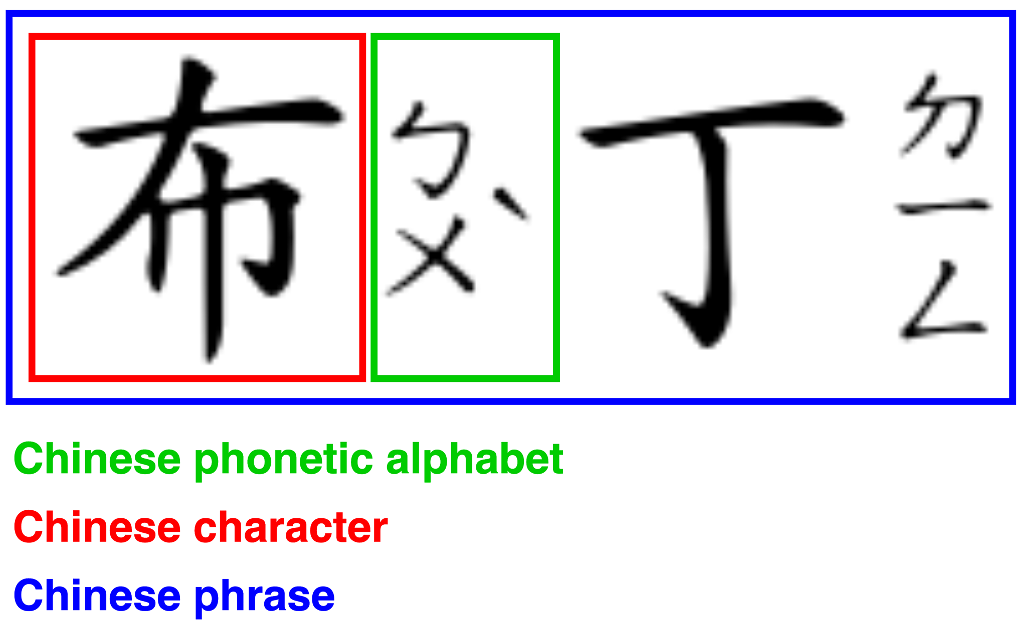}
	\caption{Definition of Chinese~words.\label{chinese-word}}
\end{figure}

Voice data were collected using a tablet computer with a microphone attached. We used rode’s smartLav microphone clipped to the subject’s clothing collar. For~this task, we programmed an app to be installed on a Samsung Galaxy Tab S3 tablet. The~microphone acquired the signal at a sampling frequency of 44.1 kHz and transmitted it to the tablet computer, and~stored it in 16-bit depth. The~database consists of 96 Chinese phrases, made up of 37 Chinese phonetic alphabets, each of which appears at the beginning, middle, and~end of the word. The~definition of Chinese words is shown in Figure~\ref{chinese-word}. The~Chinese phrases were illustrated with pictures, and~the task for the child was to name the pictures spontaneously. For~a detailed list of the Chinese phrases, please refer to Table~\ref{tab:phraseslist}. For~each recording, two SLPs prepared diagnostic notes. The~evaluation was performed to identify pathological pronunciation. In~addition, abnormal intonation sounds were analyzed, and~pathological types were annotated. Four types of articulation were collected:

\begin{itemize}
	\item	Stopping.
	\item	Backing.
	\item	FCDP.
	\item	Affrication.
\end{itemize}

Before collecting the corpus, we expected a single model to identify the corresponding error category based on the phonetic sound of a single word. When compared to other languages, Chinese SSDs have more then 15 different error categories. Still, statistical analysis of the corpus we collected revealed that four or five of them are more common in clinical cases, implying that the other categories are relatively rare. Because~it is challenging to train deep learning with extremely unbalanced or irregular data categories, we discussed with the SLPs. We determined that it would be better to start with the most common types in the clinical~setting.

Two different kinds of speech samples were used: complete Chinese phrases and single Chinese characters. Each recorded speech sample contains a complete Chinese phrase, and~a single Chinese character sample is extracted from each phrase sample. Following are the justifications for this action: The phrases are designed to use the 37 Chinese phonetic alphabets, arranging each phonetic symbol to appear in the front, middle, and~back of common Chinese phrases. It is possible for patients to make SSDs in different positions when pronouncing a word or for various positions to contain different types of SSDs. In~the case of the marker samples, only a single type of SSD is indicated in the SSD label of Chinese phrases. The~marker data do not contain possible locations and multiple classes. To~solve these problems, we designed Experiment 2 and recreated the~dataset. 

All samples were re-syllabified by acoustic experts, and~two SLPs rigorously labeled all Chinese-single-character samples. To increase the accuracy and reliability of SLPs diagnostic results, only samples with consistent SLPs labels were preserved for subsequent studies. The~SLPs labeling program used our custom-built labeling software to listen to each audio file and click on the error category option to which the sample belonged, after~which the software generated a labeled~file.

\subsection{Data~Pre-Processing}
To preserve space and time information in the conversion of sound features, we chose Mel spectrograms as the feature representation. To~perform transfer learning in a standard model pre-trained with image net, we used a three-channel Mel spectrogram. On~each channel, the~MelSpectrogram was calculated using various window widths and hop lengths of \{25 ms, 10 ms\}, \{50 ms, 25 ms\}, and~\{100 ms, 50 ms\}. Different window widths and hop lengths guaranteed that each channel had varying amounts of frequency and temporal~information.

To avoid overfitting during training and to make more efficient use of the limited sample, we used a variety of standard sound augmentation methods, as~shown in the table~below:
\begin{itemize}
	\item	Increase/decrease the pitch by two semitones.
	\item	Shift the time by 10\%.
	\item	Scaling the speed by random number within $\pm 25$\%.
	\item	The input audio signal is compressed using dynamic range compression.
	\item   Increase/decrease volume by a random number of decibels in [3, 3] dB.
	\item   Random noise in the range [0, 10] dB is added (SNR).
\end{itemize}

All expansions were implemented using ffmpge~\cite{tomar2006converting} and python librosa packages~\cite{mcfee2015librosa}. After~augmentation, we had nine times more~data.

\subsection{Models\label{2.3}}

We used three standard models to solve our problems. The~following are the~models:
\begin{enumerate}
	\item	EfficientNet~\cite{tan2019efficientnet}: They use neural architecture search to create a new baseline network and scale it up to create the EfficientNet family of models, which outperform previous ConvNets in accuracy and efficiency. EfficientNet uses a new scaling method that uses a simple but highly effective compound coefficient to scale all depth/width/resolution dimensions uniformly. EfficientNet shows how to scale up MobileNets and ResNets with this method.
	\item	\textls[-15]{DenseNet~\cite{iandola2014densenet}: Dense Convolutional Network (DenseNet) is a feed-forward network that connects each layer to every other layer. The~network has L(L + 1)/2 direct connections, whereas traditional convolutional networks with L layers have L connections between each and its subsequent layers. All previous layers' feature maps are used as inputs into each layer, and~their feature maps are used as inputs into all \mbox{successive layers.}}
	\item	InceptionV3~\cite{xia2017inception}: The Inception architecture has been shown to achieve excellent performance while using a small amount of computational power. Inception network training is significantly accelerated when residual connections are used. By~a razor-thin margin, residual Inception networks outperform similarly priced Inception networks without residual connections. They present several new streamlined Inception network architectures, both residual and non-residual.
\end{enumerate}

The model's trainable parameters and size are provided in Table~\ref{model-size-params}.

\begin{table}[H]\small
	\caption{Trainable parameters and size of the~model.}
	\setlength{\tabcolsep}{7.86mm}\begin{tabular}{rrrr}
	\toprule
	\multicolumn{1}{c}{\textbf{Experiment}} & \multicolumn{1}{c}{\textbf{Model}} & \multicolumn{1}{c}{\textbf{Trainable Params}} & \multicolumn{1}{c}{\textbf{Size (mb)}}   \\ \midrule
	\multirow{3}{*}{e1} & DenseNet121    & 6,957,956  & 82    \\
	& EfficientNetB2 & 7,706,630  & 91    \\
	& InceptionV3    & 21,776,548 & 251   \\
	\multirow{3}{*}{e2} & DenseNet121    & 6,955,906  & 82    \\
	& EfficientNetB2 & 7,703,812  & 91    \\
	& InceptionV3    & 21,772,450 & 251   \\
	\multirow{3}{*}{e3} & DenseNet121    & 6,957,956  & 82    \\
	& EfficientNetB2 & 7,706,630  & 91    \\
	& InceptionV3    & 21,776,548 & 251   \\ \bottomrule
	\end{tabular}
	\label{model-size-params}
\end{table}
\unskip

\subsection{Training~Environments}

Due to the dataset's small sample size and data imbalance, we resolved the issue using class weights. We created a 5-folder cross-validation dataset for training and evaluating the model. We separated the data into training and validation at 80\%, 20\%, respectively. We configured the batch size to be 128, the~number of epochs to 15, the~training optimizer to be Adam, and~the learning rate to 0.0001. Our loss function used categorical cross-entropy in Experiments 1 and 3 and binary cross-entropy in Experiment 2. The~model with the lowest validation loss was saved as the result of each training session. Training and validation were carried out by a Keras-based TensorFlow platform (version 2.4) on Nvidia Tesla V100 with 32GB RAM. For~training the same model, the~same framework, hyperparameter settings, and~training procedures were~used.

\subsection{Experiment~Methods}
\subsubsection{Experiment 1---Multi-Class Classification Using a Single Chinese~Phrase}

In this experiment, three standard models were used to predict four error categories by entering complete Chinese phrases. First, all audio files were labeled according to the category corresponding to the diagnostic label of SLPs. Then the feature map was processed to [128, 256, 3] size according to the preprocessing method in Section~\ref{2.3}. Finally, five folders were created for cross-validation, and~the number of data is shown in Table~\ref{e1-data}.

\begin{table}[H]\small
	\caption{The amount of data on single-Chinese-phrase dataset. Training segment contains augmented~data.}
	\begin{adjustwidth}{-\extralength}{0cm}
		\setlength{\tabcolsep}{4.3mm} \begin{tabular}{rrrrrrrrr}
		\toprule
		\multirow{2}{*}{\textbf{CV}} & \multicolumn{4}{l}{\textbf{Training Segments}} & \multicolumn{4}{l}{\textbf{Test Segments}}                                                                \\
		& \textbf{FCDP}                                  & \textbf{Affrication}                       & \textbf{Backing} & \textbf{Stopping} & \textbf{FCDP} & \textbf{Affrication} & \textbf{Backing} & \textbf{Stopping} \\
		\midrule
		Fold1               & 4401                                  & 2628                              & 1332    & 9936     & 122  & 72          & 37      & 276      \\
		Fold2               & 4401                                  & 2619                              & 1332    & 9936     & 122  & 73          & 37      & 276      \\
		Fold3               & 4401                                  & 2619                              & 1332    & 9936     & 122  & 73          & 37      & 276      \\
		Fold4               & 4401                                  & 2619                              & 1332    & 9936     & 122  & 73          & 37      & 276      \\
		Fold5               & 4401                                  & 2619                              & 1332    & 9936     & 123  & 73          & 37      & 276      \\
		\bottomrule
		\end{tabular}
	\end{adjustwidth}
	\label{e1-data}
\end{table}
\unskip

\subsubsection{Experiment 2---Binary Classification Using a Single Chinese~Character}

To mark the location of the misconstructions more precisely, the~acoustic experts re-cut all Chinese phrases into individual sound files according to the Chinese characters. It means that each sample will contain only one Chinese character. The~two SLPs re-evaluated the segmented samples and selected those with more significant error characteristics to produce a single-Chinese-character dataset. To~evaluate the model's ability to discriminate among the accurate samples, the~new dataset with corresponding correctly pronounced samples was used to test the performance of the model for binary classification. Since the length of the sound sample becomes shorter after cutting, a~feature size of [128, 128, 3] in the pre-processing is sufficient to include the sample features. In~Experiment 2, the~model took  single-Chinese-character samples as input and output them as a correct category or incorrect category, and~the amount of data is shown in Tables~\ref{e2-data-1} and~\ref{e2-data-2}.

\begin{table}[H]\small
	\caption{The amount of data on the single-Chinese-character dataset for~training.}
	\begin{adjustwidth}{-\extralength}{0cm}
		\setlength{\tabcolsep}{4.63mm}\begin{tabular}{ccccccccc}
		\toprule
		\multirow{2}{*}{\textbf{CV}} & \multicolumn{2}{c}{\textbf{Backing}} & \multicolumn{2}{c}{\textbf{Stopping}} & \multicolumn{2}{c}{\textbf{Affrication}} & \multicolumn{2}{c}{\textbf{FCDP}}                                             \\
		& \textbf{Incorrect}                   & \textbf{Correct}                      & \textbf{Incorrect}                       & \textbf{Correct}                  & \textbf{Incorrect} & \textbf{Correct} & \textbf{Incorrect} & \textbf{Correct} \\
		\midrule
		Fold1               & 1125                        & 5724                         & 1728                            & 11,655                    & 2853      & 5283    & 1089      & 3636    \\
		Fold2               & 1125                        & 5724                         & 1728                            & 11,655                    & 2853      & 5283    & 1089      & 3636    \\
		Fold3               & 1125                        & 5724                         & 1728                            & 11,655                    & 2853      & 5283    & 1089      & 3636    \\
		Fold4               & 1125                        & 5724                         & 1728                            & 11,655                    & 2853      & 5283    & 1089      & 3636    \\
		Fold5               & 1116                        & 5715                         & 1728                            & 11,646                    & 2844      & 5274    & 1080      & 3636    \\
		\bottomrule
		\end{tabular}
	\end{adjustwidth}
	\label{e2-data-1}
\end{table}
\unskip

\begin{table}[H]\small
	\caption{The amount of data on the single-Chinese-character dataset for~validation.}
	\begin{adjustwidth}{-\extralength}{0cm}
		\setlength{\tabcolsep}{4.63mm} \begin{tabular}{ccccccccc}
		\toprule
		\multirow{2}{*}{\textbf{CV}} & \multicolumn{2}{c}{\textbf{Backing}} & \multicolumn{2}{c}{\textbf{Stopping}} & \multicolumn{2}{c}{\textbf{Affrication}} & \multicolumn{2}{c}{\textbf{FCDP}}                                             \\
		& \textbf{Incorrect}                   & \textbf{Correct}                      & \textbf{Incorrect}                       & \textbf{Correct}                  & \textbf{Incorrect} & \textbf{Correct} & \textbf{Incorrect} & \textbf{Correct} \\
		\midrule
		Fold1               & 31                          & 158                          & 48                              & 323                      & 79        & 146     & 30        & 101     \\
		Fold2               & 31                          & 159                          & 48                              & 324                      & 79        & 147     & 30        & 101     \\
		Fold3               & 31                          & 159                          & 48                              & 323                      & 79        & 146     & 30        & 101     \\
		Fold4               & 31                          & 159                          & 48                              & 324                      & 79        & 147     & 30        & 101     \\
		Fold5               & 32                          & 159                          & 48                              & 324                      & 80        & 147     & 31        & 101     \\
		\bottomrule
		\end{tabular}
	\end{adjustwidth}
	\label{e2-data-2}
\end{table}
\unskip

\subsubsection{Experiment 3---Multi-Class Classification Using a Single Chinese~Character}

In Experiment 3, to~further verify the ability of the model to discriminate the four types of errors in a single model, we repackaged the dataset from Experiment 2 to leave only the samples of error tones. The~sample size of the dataset is shown in Table~\ref{e3-data}. In~this experiment, the~input of the model was a single Chinese character sample and the output was four error~categories.

\begin{table}[H]\small
	\caption{The amount of data on single-Chinese-character~dataset.}
	\begin{adjustwidth}{-\extralength}{0cm}
		\setlength{\tabcolsep}{4.3mm}\begin{tabular}{ccccccccc}
		\toprule
		\multirow{2}{*}{\textbf{CV}} & \multicolumn{4}{c}{\textbf{Training Segments}} & \multicolumn{4}{c}{\textbf{Test Segments}}                                                                \\
		& \textbf{Backing}                               & \textbf{Stopping}                          & \textbf{Affrication} & \textbf{FCDP} & \textbf{Backing} & \textbf{Stopping} & \textbf{Affrication} & \textbf{FCDP} \\
		\midrule
		Fold1               & 1125                                  & 1728                              & 2853        & 1089 & 31      & 48       & 79          & 30   \\
		Fold2               & 1125                                  & 1728                              & 2853        & 1089 & 31      & 48       & 79          & 30   \\
		Fold3               & 1125                                  & 1728                              & 2853        & 1089 & 31      & 48       & 79          & 30   \\
		Fold4               & 1125                                  & 1728                              & 2853        & 1089 & 31      & 48       & 79          & 30   \\
		Fold5               & 1116                                  & 1728                              & 2844        & 1080 & 32      & 48       & 80          & 31   \\
		\bottomrule
		\end{tabular}
	\end{adjustwidth}
	\label{e3-data}
\end{table}
\unskip

\subsubsection{Runtime of the Developed~Application}
We converted the trained models into TensorFlow Lite models (.tflite), and~measured the inference time of all models on an Android mobile device. Given that the model is intended to be used in real-time by physicians or patients via smartphones, the~time taken to infer is also critical. We used the performance measurement application~\cite{Abadi_TensorFlow_Large-scale_machine_2015} provided by the official Tensorflow website for performance evaluation. We tested all the models used using Google Pixel 6 with Android 12.

\section{Results\label{3}}

To describe the experimental results, this chapter is divided into several subheadings. The~first section will look into the efficacy of using Chinese phrases as classifier input. The~second section investigates the effectiveness of Chinese characters in contrast. The~third subsection investigates the efficacy of Chinese characters in the classifier. Furthermore, real-time inference on mobile devices is provided to demonstrate the viability of edge prediction. Initially, the~unbalanced dataset led to ineffective training outcomes, which were not significantly enhanced until we implemented the balancing measures of class weights and data~augmentation.

\subsection{Experiment 1---Multi-Class Classification Using a Single Chinese~Phrase}

\begin{figure}[H]
	\includegraphics[width=10 cm]{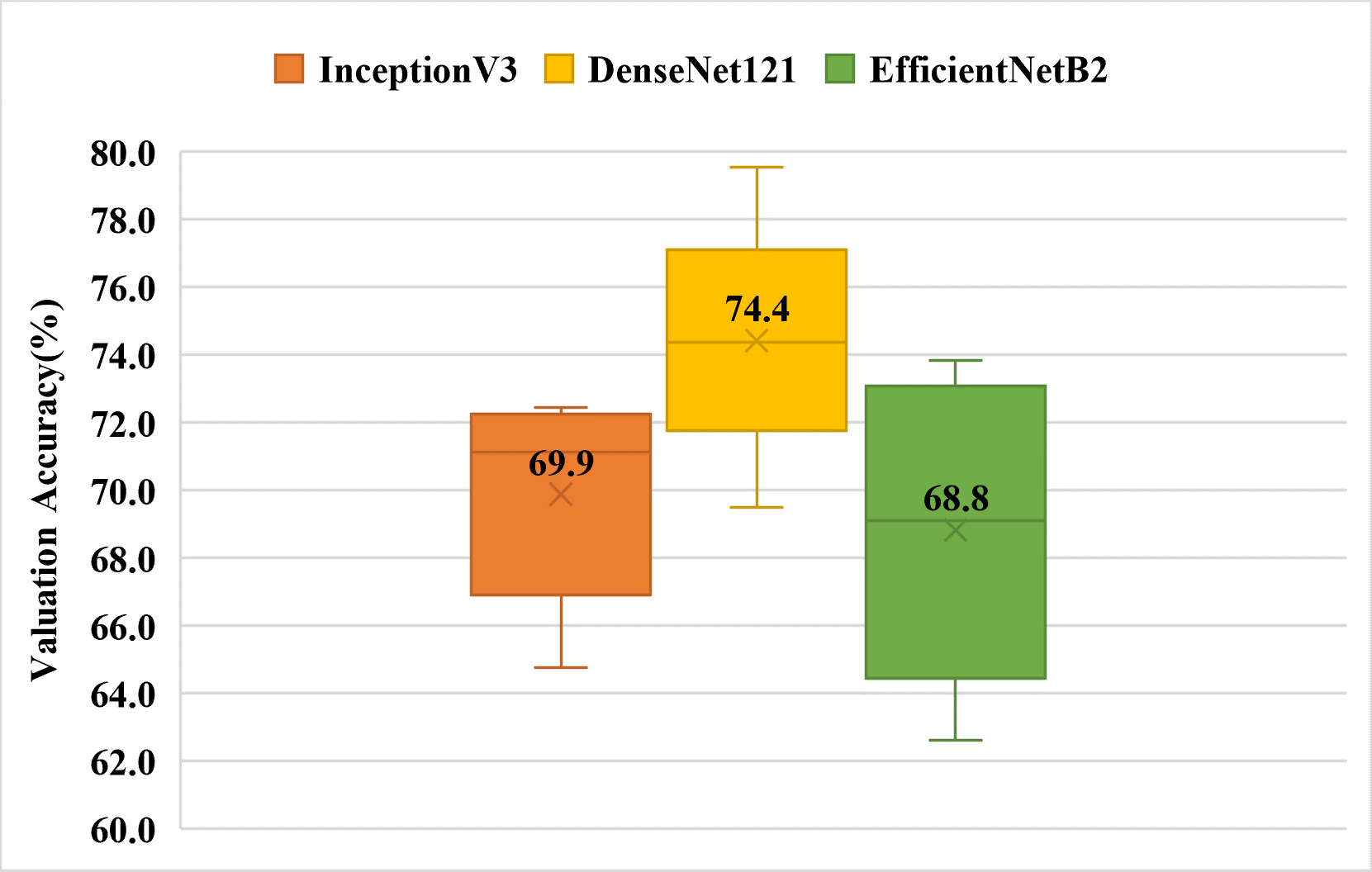}
	\caption{Validation accuracy of single Chinese phrase multi-category classification box plot (Experiment-1). The~top and bottom of the box are the interquartile ranges (75th and 25th~percentile) centered around the median value (50th~percentile). The~whiskers represent the minimum and maximum validation accuracy values. Table~\ref{tab:e1} presents the results in~detail.}
	\label{fig2}
\end{figure}

\textls[-15]{To verify the feasibility of the model for classifying SSDs, we performed cross-training on three standard models. Figure~\ref{fig2} shows the training results of the dataset on each model. The~average cross-validation results for the three models were as follows: InceptionV3 with a result of 70\%, DenseNet121 with a result of 74\%, and~EfficientNetB2 with a result of 69\%. Table~\ref{e1_cm} shows the confusion matrix with the best accuracy among all the~results.}

\unskip

\begin{table}[H]\small
	\caption{One of the most effective confusion matrices in Experiment 1 when the DenseNet121 model was used. Rows indicate output classes, columns indicate target~classes.}
	\setlength{\tabcolsep}{7.37mm}   \begin{tabular}{rrrrr}
	\toprule
	& \textbf{FCDP}         & \textbf{Affrication} & \textbf{Backing}     & \textbf{Stopping}     \\
	\midrule
	FCDP        & 120  & 0           & 0           & 2            \\
	Affrication & 4            & 65 & 0           & 4            \\
	Backing     & 7            & 2           & 20 & 8            \\
	Stopping    & 8            & 3           & 3           & 262 \\
	\bottomrule
	\end{tabular}
	\label{e1_cm}
\end{table}
\unskip

\begin{table}[H]\small
	\caption{Validation accuracy of single Chinese phrase multi-category~classification.}
	\setlength{\tabcolsep}{7.52mm}\begin{tabular}{rrrr}
	\toprule
	\textbf{Fold   Number} & \textbf{InceptionV3} & \textbf{DenseNet121} & \textbf{EfficientNetB2} \\ \midrule
	1             & 69.0        & 74.4        & 66.3           \\
	2             & 72.1        & 74.0        & 69.1           \\
	3             & 72.4        & 79.5        & 73.8           \\
	4             & 64.8        & 69.5        & 62.6           \\
	5             & 71.1        & 74.7        & 72.3           \\
	Average Value & 69.9        & 74.4        & 68.8           \\ \bottomrule
	\end{tabular}
	\label{tab:e1}
\end{table}
\unskip

\subsection{Experiment 2---Binary Classification Using a Single Chinese~Character}

\begin{figure}[H]
	\includegraphics[width=9 cm]{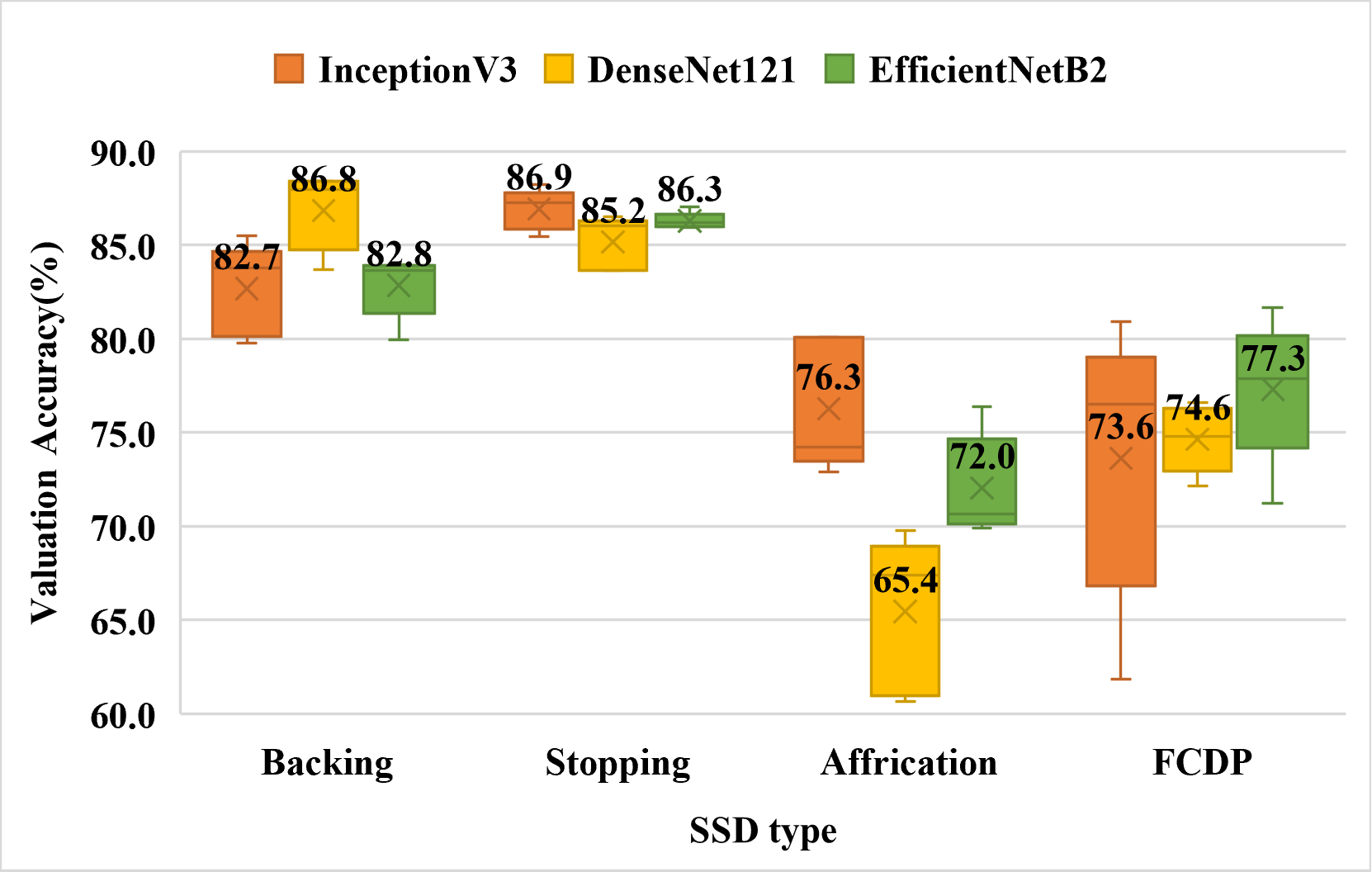}
	\caption{Validation accuracy of single Chinese character binary classification box plot (Experiment-2). Tables~\ref{tab:e2-Backing}--\ref{tab:e2-FCDP} present the results in~detail.}
	\label{fig3}
\end{figure}

Figure~\ref{fig3} shows the accuracy results of the four types of binary classifications with phonetic errors on the three models. The~displayed numbers are the average of the cross-validation results, and~the following are the best results in each category: backing is 86.8~percent of DenseNet121; stopping is 86.9~percent of InceptionV3; Affricate is 76.3~percent of InceptionV3; and FCDP is 76~percent of EffcientNetB2. It can be found that the results of affrication and FCDP in each folder are relatively different, which we speculate is due to the fact that these two categories contain more Chinese characters, and the number of samples currently collected is not enough to satisfy the plurality of~data.

\unskip

\begin{table}[H]\small
	\caption{Validation accuracy of single Chinese character binary classification of backing~class.}
	\setlength{\tabcolsep}{7.52mm}\begin{tabular}{rlll}
	\toprule
	\textbf{Fold   Number} & \multicolumn{1}{r}{\textbf{InceptionV3}} & \multicolumn{1}{r}{\textbf{DenseNet121}} & \multicolumn{1}{r}{\textbf{EfficientNetB2}} \\ \midrule
	1             & 83.8                            & 88.4                            & 79.9                               \\
	2             & 79.8                            & 83.7                            & 83.9                               \\
	3             & 83.8                            & 85.8                            & 84.0                               \\
	4             & 80.5                            & 88.4                            & 82.8                               \\
	5             & 85.5                            & 88.0                            & 83.7                               \\
	Average Value & 82.7                            & 86.8                            & 82.8                               \\ \bottomrule
	\end{tabular}
	\label{tab:e2-Backing}
\end{table}
\unskip

\begin{table}[H]\small
	\caption{\textls[-15]{Validation accuracy of single Chinese character multi-category classification of stopping~class}.}
	\label{tab:e2-Stopping}
	\setlength{\tabcolsep}{7.52mm}\begin{tabular}{rlll}
	\toprule
	\textbf{Fold   Number} & \multicolumn{1}{r}{\textbf{InceptionV3}} & \multicolumn{1}{r}{\textbf{DenseNet121}} & \multicolumn{1}{r}{\textbf{EfficientNetB2}} \\ \midrule
	1             & 87.3                            & 86.5                            & 86.2                               \\
	2             & 88.2                            & 86.1                            & 86.0                               \\
	3             & 87.4                            & 83.7                            & 86.3                               \\
	4             & 85.4                            & 86.0                            & 86.0                               \\
	5             & 86.3                            & 83.7                            & 87.1                               \\
	Average Value & 86.9                            & 85.2                            & 86.3                               \\ \bottomrule
	\end{tabular}
\end{table}
\unskip

\begin{table}[H]\small
	\caption{\textls[-15]{Validation accuracy of single Chinese character multi-category classification of affrication~class.}}
	\label{tab:e2-Affrication}
	\setlength{\tabcolsep}{7.52mm}\begin{tabular}{rlll}
	\toprule
	\textbf{Fold   Number} & \multicolumn{1}{r}{\textbf{InceptionV3}} & \multicolumn{1}{r}{\textbf{DenseNet121}} & \multicolumn{1}{r}{\textbf{EfficientNetB2}} \\ \midrule
	1             & 80.1                            & 69.8                            & 69.9                               \\
	2             & 74.0                            & 60.7                            & 70.7                               \\
	3             & 74.2                            & 67.4                            & 72.9                               \\
	4             & 72.9                            & 61.3                            & 76.4                               \\
	5             & 80.1                            & 68.1                            & 70.4                               \\
	Average Value & 76.3                            & 65.4                            & 72.0                               \\ \bottomrule
	\end{tabular}
\end{table}
\unskip

\begin{table}[H]\small
	\caption{Validation accuracy of single Chinese character multi-category classification of FCDP~class.}
	\label{tab:e2-FCDP}
	\setlength{\tabcolsep}{7.52mm}\begin{tabular}{rlll}
	\toprule
	\textbf{Fold   Number} & \multicolumn{1}{r}{\textbf{InceptionV3}} & \multicolumn{1}{r}{\textbf{DenseNet121}} & \multicolumn{1}{r}{\textbf{EfficientNetB2}} \\ \midrule
	1             & 71.8                            & 74.8                            & 77.9                               \\
	2             & 76.5                            & 76.6                            & 77.1                               \\
	3             & 61.8                            & 73.7                            & 81.7                               \\
	4             & 80.9                            & 72.2                            & 78.6                               \\
	5             & 77.1                            & 76.0                            & 71.2                               \\
	Average Value & 73.6                            & 74.6                            & 77.3                               \\ \bottomrule
	\end{tabular}
\end{table}
\unskip

\subsection{Experiment 3---Multi-Class Classification Using a Single Chinese~Character}

\begin{figure}[H]
	\includegraphics[width=10 cm]{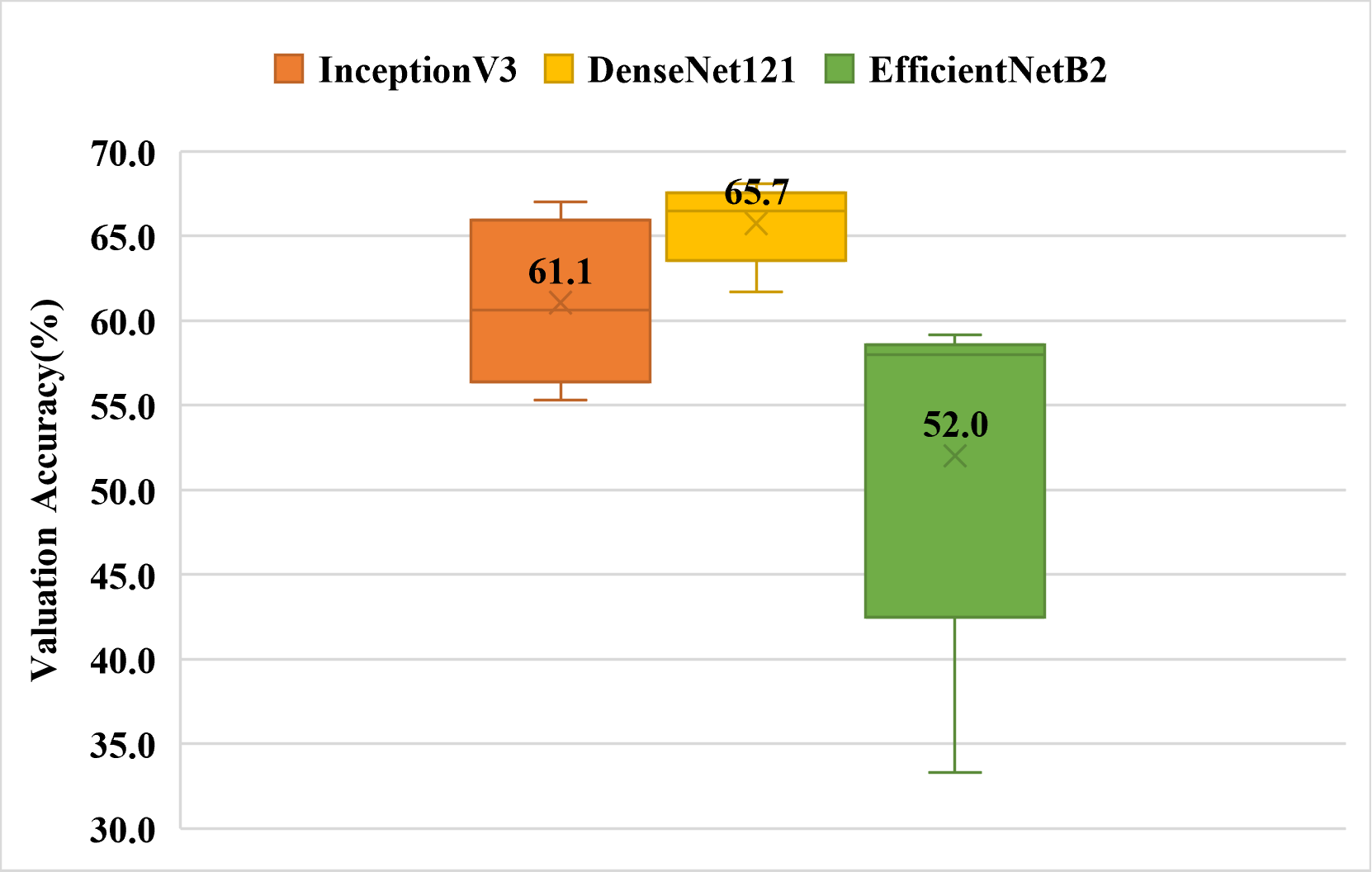}
	\caption{Validation accuracy of single Chinese character multi-category classification box plot (Experiment-3). Table~\ref{tab:e3} presents the results in~detail.}
	\label{fig4}
\end{figure}

The experimental results are shown in Figure~\ref{fig4}. It can be found that the overall accuracy of the model has decreased somewhat compared with that of Experiment 1, but~the confusion matrix in Table~\ref{e3_cm} shows that the model still has a certain level of discriminatory~ability.

\unskip

\begin{table}[H]\small
	\caption{One of the most effective confusion matrices in Experiment 3 when the InceptionV3 model was used. Rows indicate output classes, columns indicate target~classes.}
	\setlength{\tabcolsep}{7.38mm}   \begin{tabular}{ccccc}
	\toprule
	& \textbf{Backing}     & \textbf{Stopping}    & \textbf{Affrication} & \textbf{FCDP}        \\ \midrule
	Backing     & 19  & 3           & 5           & 5           \\
	Stopping    & 7           & 27 & 11          & 3           \\
	Affrication & 1           & 10          & 65 & 2           \\
	FCDP        & 2           & 0           & 7           & 22 \\ \bottomrule
	\end{tabular}
	\label{e3_cm}
\end{table}
\unskip

\begin{table}[H]\small
	\caption{Validation accuracy of single Chinese character multi-category~classification.}
	\label{tab:e3}
	\setlength{\tabcolsep}{7.52mm}\begin{tabular}{rlll}
	\toprule
	\textbf{Fold   Number} & \multicolumn{1}{r}{\textbf{InceptionV3}} & \multicolumn{1}{r}{\textbf{DenseNet121}} & \multicolumn{1}{r}{\textbf{EfficientNetB2}} \\ \midrule
	1             & 55.3                            & 65.4                            & 51.6                               \\
	2             & 60.6                            & 61.7                            & 58.0                               \\
	3             & 64.9                            & 66.5                            & 58.0                               \\
	4             & 57.5                            & 67.0                            & 59.2                               \\
	5             & 67.0                            & 68.1                            & 33.3                               \\
	Average Value & 61.1                            & 65.7                            & 52.0                               \\ \bottomrule
	\end{tabular}
\end{table}
\unskip

\subsection{Runtime of the Developed~Application}

The performance of all the models is summarized in Table~\ref{tflite}, and~it is clear that the models we used meet the requirements for real-world usage scenarios. In~other words, it only takes about three seconds for the GPU to predict all 96 Chinese phrases on the phone. The~accuracy of the TFLite model run on a cell phone was nearly identical (less than one~percent) to that of the original model run on a~computer.

\begin{table}[H]\small
	\caption{The performance values below are measured on Android 12 from Google Pixel~6.}
	\label{tflite}
	\setlength{\tabcolsep}{5.86mm} \begin{tabular}{lrrrr}
	\toprule
	\textbf{Experiment}          & \textbf{Model Name}      & \textbf{CPU}    & \textbf{GPU}   & \textbf{Model Size (MB)} \\
	\midrule
	\multirow{3}{*}{e1} & DenseNet 121    & 489 ms & 32 ms & 27             \\
	& EfficientNet B2 & 438 ms & -   & 29             \\
	& Inception V3    & 399 ms & 35 ms & 83             \\
	\midrule
	\multirow{3}{*}{e2} & DenseNet 121    & 231 ms & 29 ms & 27             \\
	& EfficientNet B2 & 241 ms & -   & 29             \\
	& Inception V3    & 170 ms & 36 ms & 83             \\
	\midrule
	\multirow{3}{*}{e3} & DenseNet 121    & 236 ms & 30 ms & 27             \\
	& EfficientNet B2 & 235 ms & -   & 29             \\
	& Inception V3    & 171 ms & 36 ms & 83             \\
	\bottomrule
	\end{tabular}
\end{table}
\unskip

\section{Discussion\label{4}}

The research presented in this article aims to develop a tool for analyzing SSD error classes based on deep learning. A~workflow has been created to collect and train a model that categorizes SSDs. SLPs, who will be the primary beneficiaries, were involved in every aspect of the study. Experts tagged the data, then analyzed and experimented with it to train the model to detect and classify SSDs. The~system is designed to help preschool children because diagnosis and intervention are most beneficial at this~age.

The results show that the use of Chinese phrase samples for the current dataset is more effective than single Chinese character samples for model training. In~general, the~four types of error categories using either Chinese phrases or single Chinese characters can achieve good results in the current mainstream image classification neural networks. However, using Chinese phrase samples as model input is easier to train than single Chinese characters samples, contrary to the original expectation. Before~the experiment, we hypothesized that reducing the range of speech marks would make it easier for the model to distinguish SSD~classes.

Several factors may account for this, including the re-screening of all samples in the Chinese single-character dataset and the elimination of ambiguous or imprecise phonetic samples by SLPs. This reduced the number of samples in the dataset. Another possible reason is that the reduced sample length also means that the model cannot find the position of the Chinese character in the original vocabulary and the combination or variation with the preceding and following sounds. This may require further refinement of the tagging method and model design to verify whether the Chinese phrase or the Chinese character is more suitable for the composition of the SSDs classification~input.

Experiments reveal that when all three models are trained under the identical conditions, the~best achievable accuracy is comparable. However, the~disparity between individual cross-training results is enormous. We believe this may have something to do with the size of the dataset. The~corpus that we have compiled must continue to be expanded so that the model can completely learn the diversity of data and more specifics during the learning process. Based on the existing training environment, all three models can effectively train usable outcomes, but~if we wish to further enhance the accuracy, we must increase the size of the voice~samples.

\section{Conclusions\label{5}}
In this paper, we investigated the idea of using neural networks to classify SSD categories in both binary and multi-category classifications. The~task is to identify the error category by the pronunciation of common Chinese words. The~categories include stopping, backing, FCDP, and~affrication. With~the progressive development of multi-dimensional CNN models, we used several standard models which are well established and powerful in image classification tasks for identification and classification. We used multi-dimensional spectral signals as input to the model, and~the input features are composed of three two-dimensional Mel-Spectrogram feature~maps.

We were able to classify four common types of SSD errors using monosyllabic speech samples and neural network models. This study is the first in Taiwan to apply deep learning to the treatment of SSDs, and~its findings are based on the four most common articulation errors in Taiwan. Possibly in the near future, machine learning will be able to aid SLP and the patient's treatment process. We found that with sufficient data, the~neural network model is able to identify subtle differences in the characteristics of different prosodic errors in single Chinese characters. Other rare categories, in~theory, can be successfully identified if sufficient samples of speech sounds are~collected.

Currently, we are converting the trained models into models that can be predicted on smartphones in a timely manner through tensorflow lite. The~pre-developed app provides a complete experience of real-time recording and analysis and~is being clinically tested in the rehabilitation department of a hospital. The~demo of the application is shown in Figure~\ref{Screenshot}. The~current accuracy of 86\% is sufficient for rapid screening for parents prior to medical treatment or self-assessment for long-term review and correction. This will save many patients or SLPs a lot of~time.



\vspace{6pt}



\authorcontributions{Conceptualization, Y.-M.K.; methodology, Y.-M.K.; software, Y.-M.K.; validation, Y.-M.K.; formal analysis, Y.-M.K.; investigation, Y.-M.K. and S.-J.R.; resources, Y.-W.T. and Y.-C.C.; data curation, Y.-M.K.; writing---original draft preparation, Y.-M.K.; writing---review and editing, S.-J.R.; visualization, Y.-M.K.; supervision, S.-J.R., Y.-W.T. and Y.-C.C.; project administration, Y.-M.K.; funding acquisition, S.-J.R., Y.-W.T. and Y.-C.C. All authors have read and agreed to the published version of the~manuscript.}

\funding{This research was funded by the National Taiwan University of Science and Technology—Cathay General Hospital Joint Research Program under the project “Classification of Articulation Disorders base on Deep Learning” (Grant number CGH-107073).}

\institutionalreview{The study was conducted according to the guidelines of the Declaration of Helsinki and approved by the Institutional Review Board of Cathay General Hospital (protocol code CGH-P107073 and June the twenty-second, 2019 of approval).}

\informedconsent{Informed consent was obtained from all subjects involved in the~study.}

\dataavailability{Not applicable.}

\acknowledgments{We thank to National Center for High-performance Computing (NCHC) for providing computational and storage resources.}

\conflictsofinterest{The authors declare no conflict of~interest.}



\abbreviations{Abbreviations}{
	The following abbreviations are used in this manuscript:\\
	
	\noindent
	\begin{tabular}{@{}ll}
		SSDs & Speech sound disorders              \\
		SLPs & Speech-language pathologists        \\
		MFCC & Mel-frequency cepstral coefficients \\
		FCDP & Final consonant deletion process    \\
	\end{tabular}}

\appendixtitles{no} 
\appendixstart
\appendix
\section[\appendixname~\thesection]{}
Table~\ref{tab:phraseslist} shows the list of Chinese phrases collected in this experiment. Each participant's pronunciation sample was recorded according to the phrases in the list. On~average, it took about 30 min to record one participant. The~length of each Chinese phrase was limited to less than three~seconds.

\begin{CJK*}{UTF8}{bsmi}
	
	\begin{table}[H]\small
		\caption{Chinese phrase~list.}
		\label{tab:phraseslist}
		\begin{adjustwidth}{-\extralength}{0cm}
			\setlength{\tabcolsep}{3.07mm} \begin{tabular}{llllll}
			\toprule
			\textbf{Chinese Phease} & \textbf{IPA}          & \textbf{Translation in English} & \textbf{Chinese Phease} & \textbf{IPA}          & \textbf{Translation in English} \\
			\midrule
			布丁           & Bùdīng       & pudding                & 閃電           & shǎndiàn     & lightning              \\
			麵包           & miànbāo      & bread                  & 牙刷           & yáshuā       & toothbrush             \\
			大白菜         & dàbáicài     & Chinese cabbage        & 直升機         & zhíshēngjī   & helicopter             \\
			螃蟹           & pángxiè      & Crab                   & 日歷           & rìlì         & calendar               \\
			奶瓶           & nǎipíng      & baby bottle            & 超人           & chāorén      & superman               \\
			蓮蓬頭         & liánpengtóu  & shower head            & 大榕樹         & dàróngshù    & Large banyan           \\
			帽子           & màozi        & hat                    & 走路           & zǒulù        & walk                   \\
			玉米           & yùmǐ         & corn                   & 洗澡           & xǐzǎo        & bath                   \\
			捉迷藏         & zhuōmícáng   & hide and seek          & 水族箱         & shuǐzúxiāng  & aquarium               \\
			鳳梨           & fènglí       & pineapple              & 草莓           & cǎoméi       & Strawberry             \\
			衣服           & yīfú         & clothing               & 洋蔥           & yángcōng     & onion                  \\
			吹風機         & chuīfēngjī   & hair dryer             & 上廁所         & shàngcèsuǒ   & To the restroom        \\
			動物           & dòngwù       & animal                 & 掃把           & sàobǎ        & broom                  \\
			蝴蝶           & húdié        & Butterfly              & 垃圾           & lèsè         & Rubbish                \\
			看電視         & kàndiànshì   & watch TV               & 去散步         & qùsànbù      & go for a walk          \\
			太陽           & tàiyáng      & Sun                    & 衣服           & yīfú         & clothing               \\
			枕頭           & zhěntou      & Pillow                 & 果醬           & guǒjiàng     & jam                    \\
			一條魚         & yītiáoyú     & a fish                 & 指甲刀         & zhǐjiǎdāo    & nail clippers          \\
			鈕扣           & niǔkòu       & button                 & 筷子           & kuàizi       & Chopsticks             \\
			電腦           & diànnǎo      & computer               & 烏龜           & wūguī        & tortoise               \\
			喝奶昔         & hēnǎixī      & drink milkshake        & 去公園         & qùgōngyuán   & go to the park         \\
			老虎           & lǎohǔ        & Tiger                  & 杜鵑花         & dùjuānhuā    & Rhododendron           \\
			恐龍           & kǒnglóng     & Dinosaur               & 選擇           & xuǎnzé       & choose                 \\
			養樂多         & yǎnglèduō    & Yakult                 & 缺點           & quēdiǎn      & shortcoming            \\
			果凍           & guǒdòng      & jelly                  & 太陽           & tàiyáng      & Sun                    \\
			烏龜           & wūguī        & tortoise               & 大海           & dàhǎi        & the sea                \\
			去公園         & qùgōngyuán   & go to the park         & 喝奶昔         & hēnǎixī      & drink milkshake        \\
			筷子           & kuàizi       & Chopsticks             & 草莓           & cǎoméi       & Strawberry             \\
			貝殼           & bèiké        & shell                  & 貝殼           & bèiké        & shell                  \\
			巧克力         & qiǎokèlì     & chocolate              & 水族箱         & shuǐzúxiāng  & aquarium               \\
			漢堡           & hànbǎo       & hamburger              & 帽子           & màozi        & hat                    \\
			大海           & dàhǎi        & the sea                & 麵包           & miànbāo      & bread                  \\
			救護車         & jiùhùchē     & ambulance              & 一條魚         & yītiáoyú     & a fish                 \\
			膠帶           & jiāodài      & adhesive tape          & 鈕扣           & niǔkòu       & button                 \\
			果醬           & guǒjiàng     & jam                    & 枕頭           & zhěntou      & Pillow                 \\
			指甲刀         & zhǐjiǎdāo    & nail clippers          & 中秋節         & zhōngqiūjié  & Mid-Autumn Festival    \\
			鉛筆           & qiānbǐ       & pencil                 & 漢堡           & hànbǎo       & hamburger              \\
			鋼琴           & gāngqín      & piano                  & 電腦           & diànnǎo      & computer               \\
			中秋節         & zhōngqiūjié  & Mid-Autumn Festival    & 看電視         & kàndiànshì   & watch TV               \\
			信封           & xìnfēng      & envelope               & 信封           & xìnfēng      & envelope               \\
			點心           & diǎnxīn      & dessert                & 鋼琴           & gāngqín      & piano                  \\
			口香糖         & kǒuxiāngtáng & chewing gum            & 吃點心         & chīdiǎnxīn   & eat dessert            \\
			站牌           & zhànpái      & stop sign              & 螃蟹           & pángxiè      & Crab                   \\
			蠟燭           & làzhú        & Candle                 & 果醬           & guǒjiàng     & jam                    \\
			擦桌子         & cāzhuōzi     & wipe the table         & 口香糖         & kǒuxiāngtáng & chewing gum            \\
			抽屜           & chōutì       & drawer                 & 鳳梨           & fènglí       & pineapple              \\
			警察           & jǐngchá      & Policemen              & 奶瓶           & nǎipíng      & baby bottle            \\
			柳橙汁         & liǔchéngzhī  & orange juice           & 蓮蓬頭         & liánpengtóu  & shower head            \\
			\bottomrule
			\end{tabular}
		\end{adjustwidth}
	\end{table}
\end{CJK*}

\section[\appendixname~\thesection]{}
The following Figure~\ref{Screenshot} shows the final mobile application, which can be used by users to perform real-time testing of the SSD category on their cell phones. The~application is divided into three main functions. First, users can download the latest version of the SSD identification model from the first screen. Then, after~filling out the basic questionnaire, the~user can enter the Chinese phrase recording stage, and~after each phrase is recorded, the~user can listen to it again and again to confirm that it is completely recorded. The~program will calculate the prediction results for each phrase and present them to the user in a report. Users can also choose whether or not to provide the recording data for backend analysis and addition to the~dataset.

\begin{figure}[H]
	\begin{adjustwidth}{-\extralength}{0cm}
		\centering 
		\begin{tabular}{@{}c@{}}
			\includegraphics[width=3.3cm]{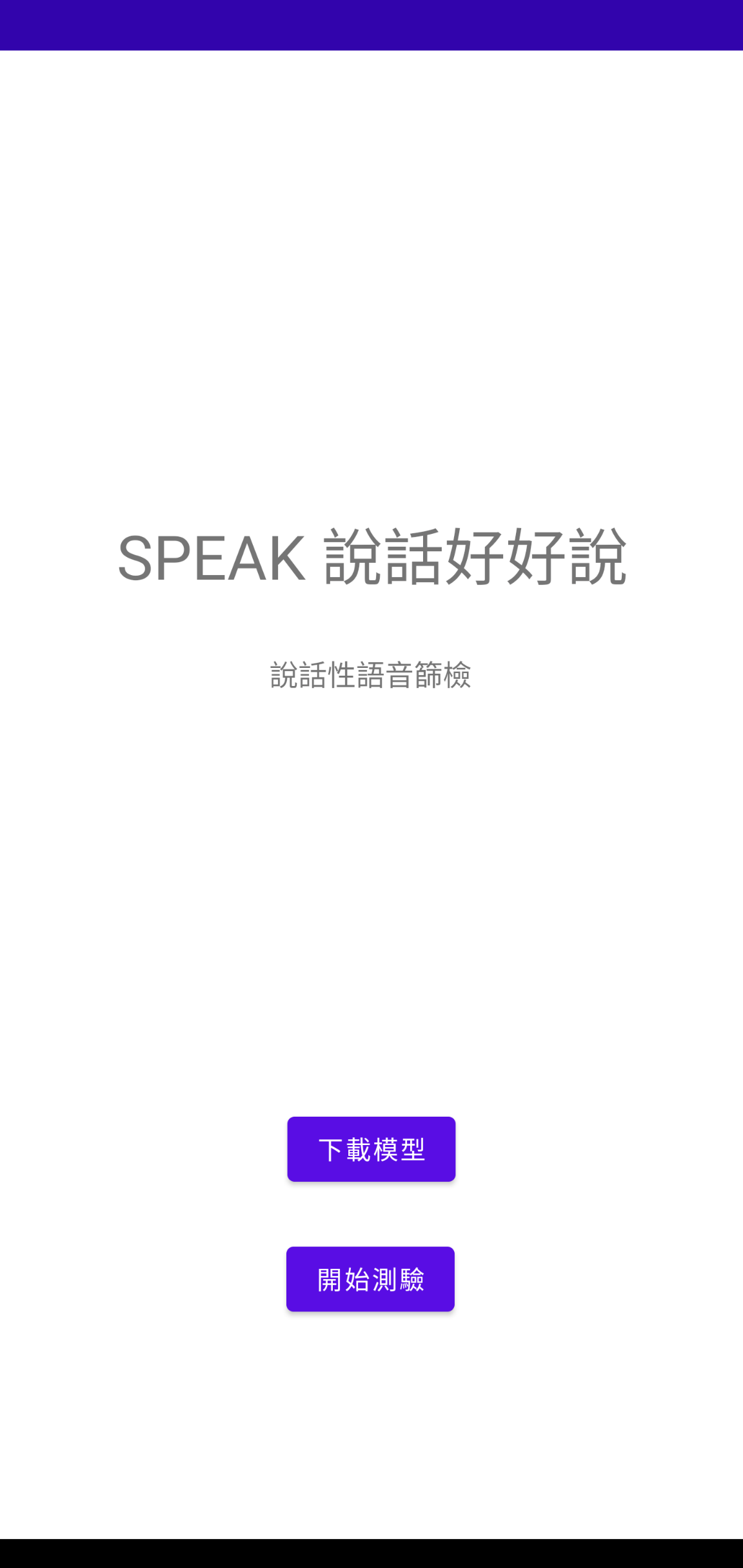} \\
			\small (\textbf{a})                     
			\label{fig:1}                           
		\end{tabular}\hfil 
		\begin{tabular}{@{}c@{}}
			\includegraphics[width=3.3cm]{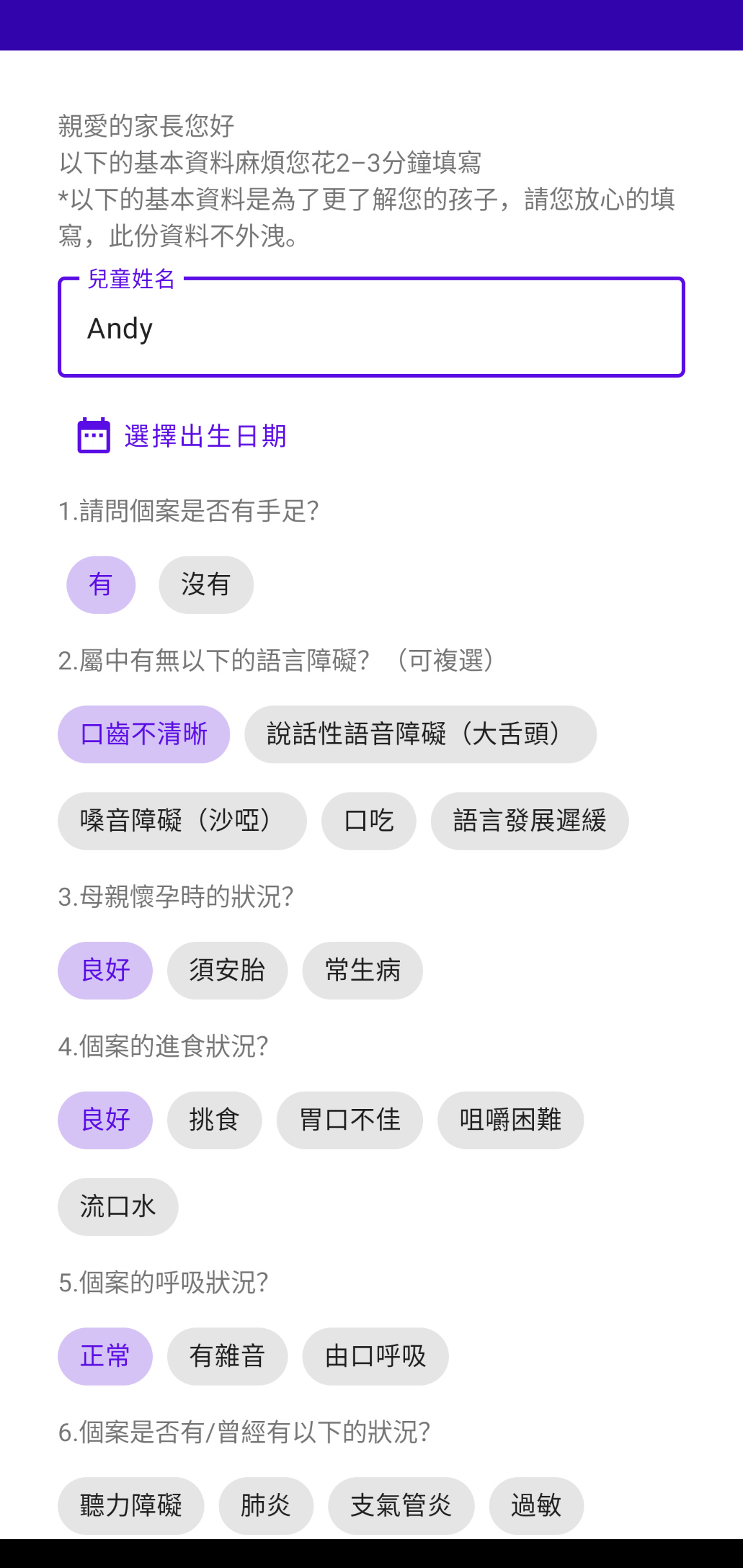} \\
			\small (\textbf{b})                     
			\label{fig:2}                           
		\end{tabular}\hfil 
		\begin{tabular}{@{}c@{}}
			\includegraphics[width=3.3cm]{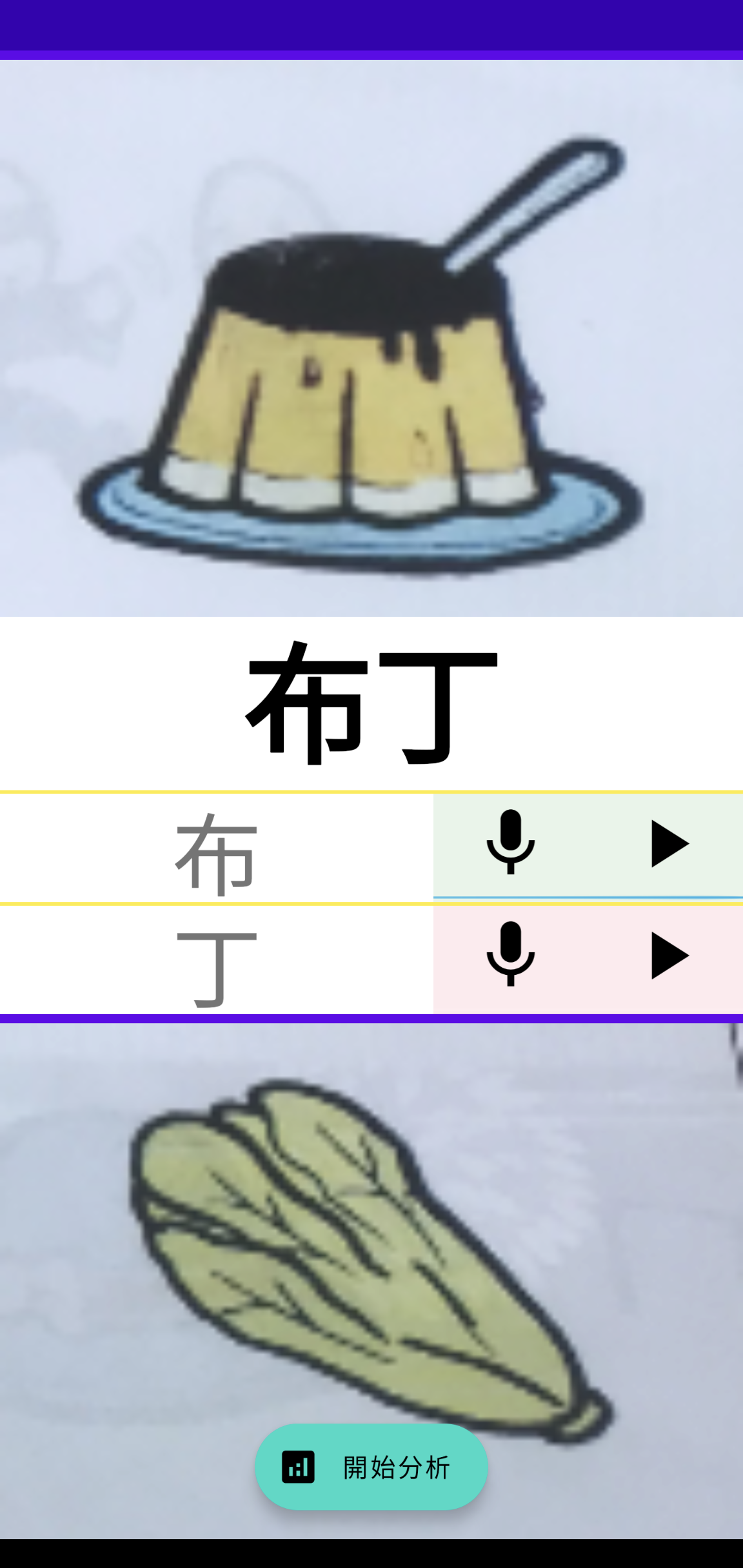} \\
			\small (\textbf{c})                     
			\label{fig:3}                           
		\end{tabular}
		
		\medskip
		\begin{tabular}{@{}c@{}}
			\includegraphics[width=3.3cm]{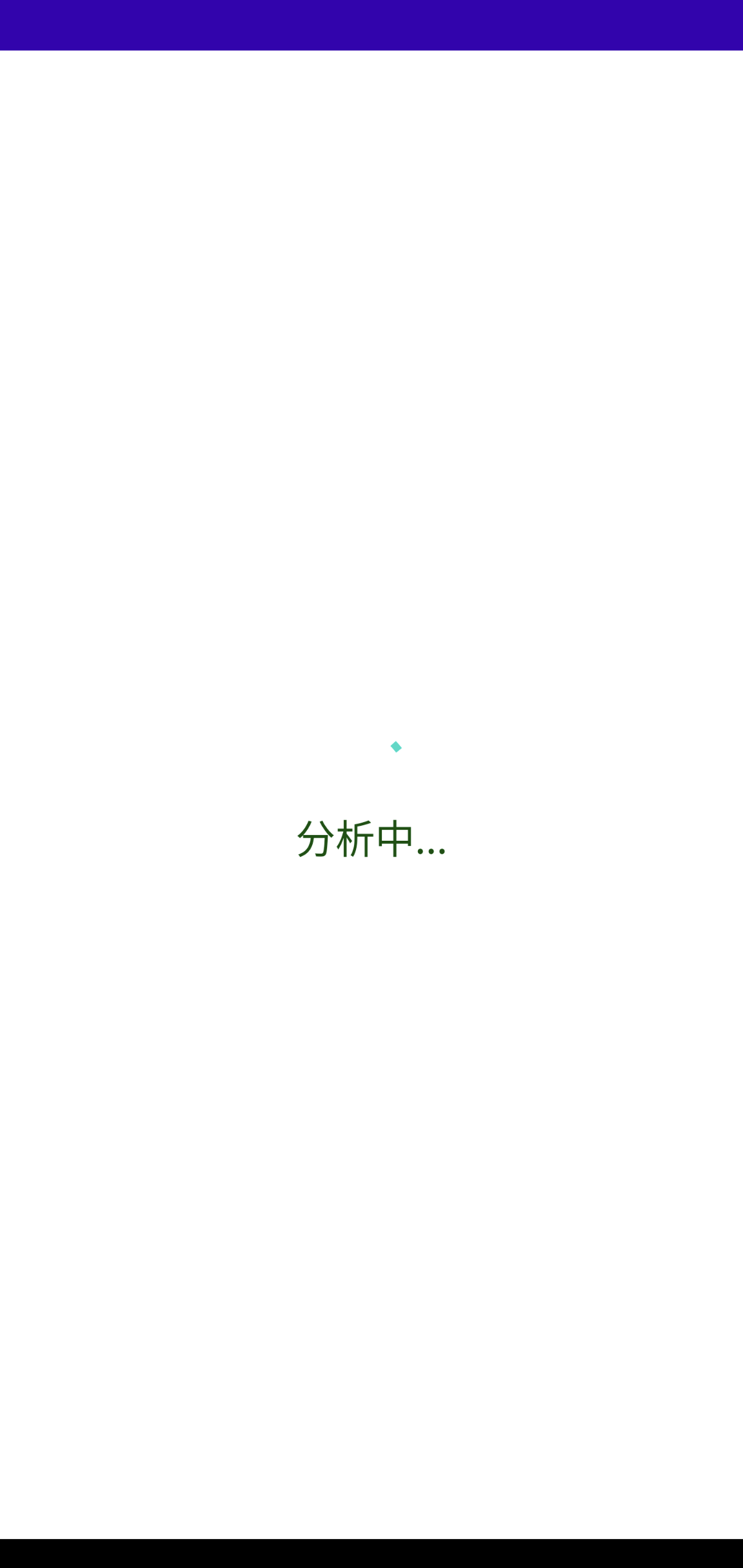} \\
			\small (\textbf{d})                     
			\label{fig:4}                           
		\end{tabular}\hfil 
		\begin{tabular}{@{}c@{}}
			\includegraphics[width=3.3cm]{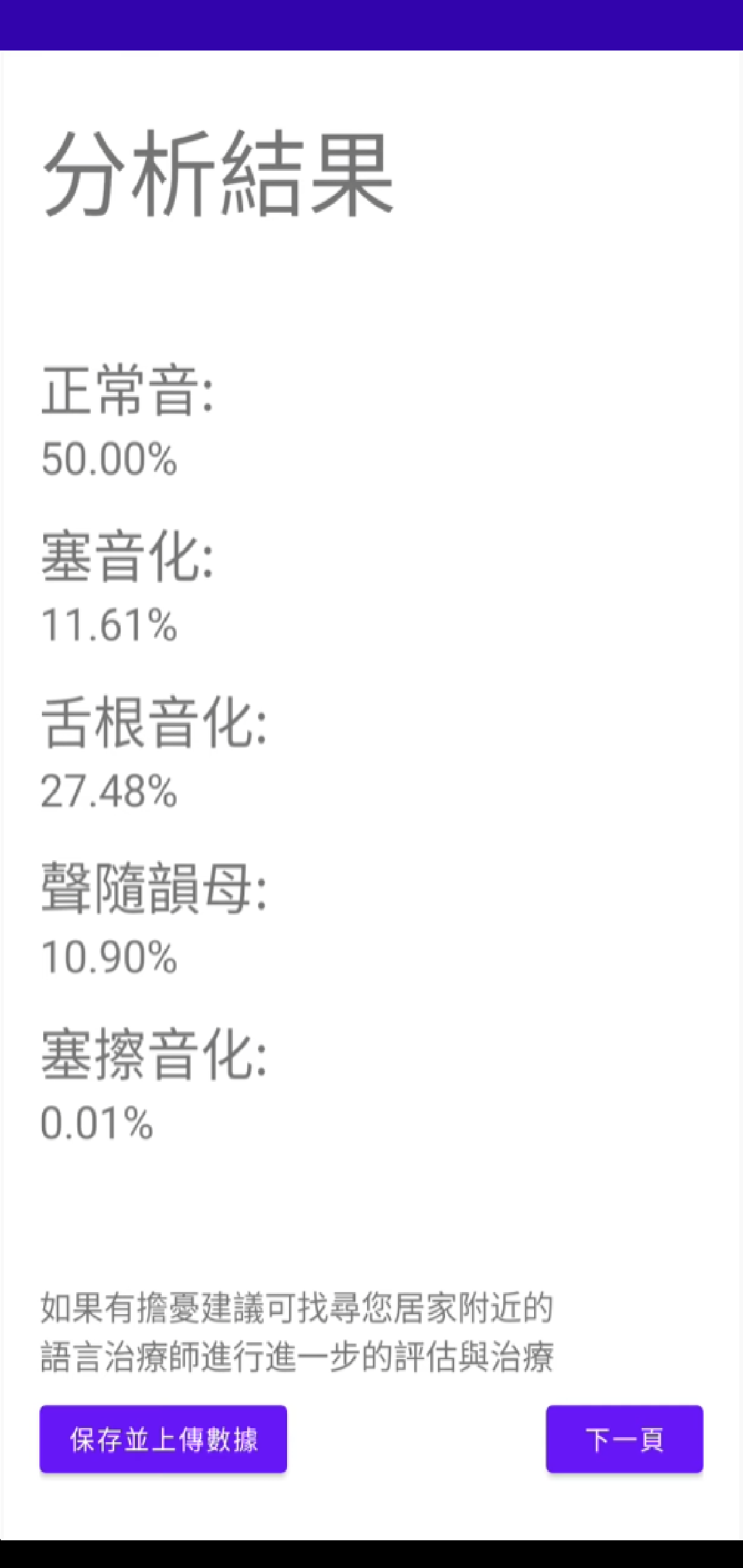} \\
			\small (\textbf{e})                     
			\label{fig:5}                           
		\end{tabular}\hfil 
		\begin{tabular}{@{}c@{}}
			\includegraphics[width=3.3cm]{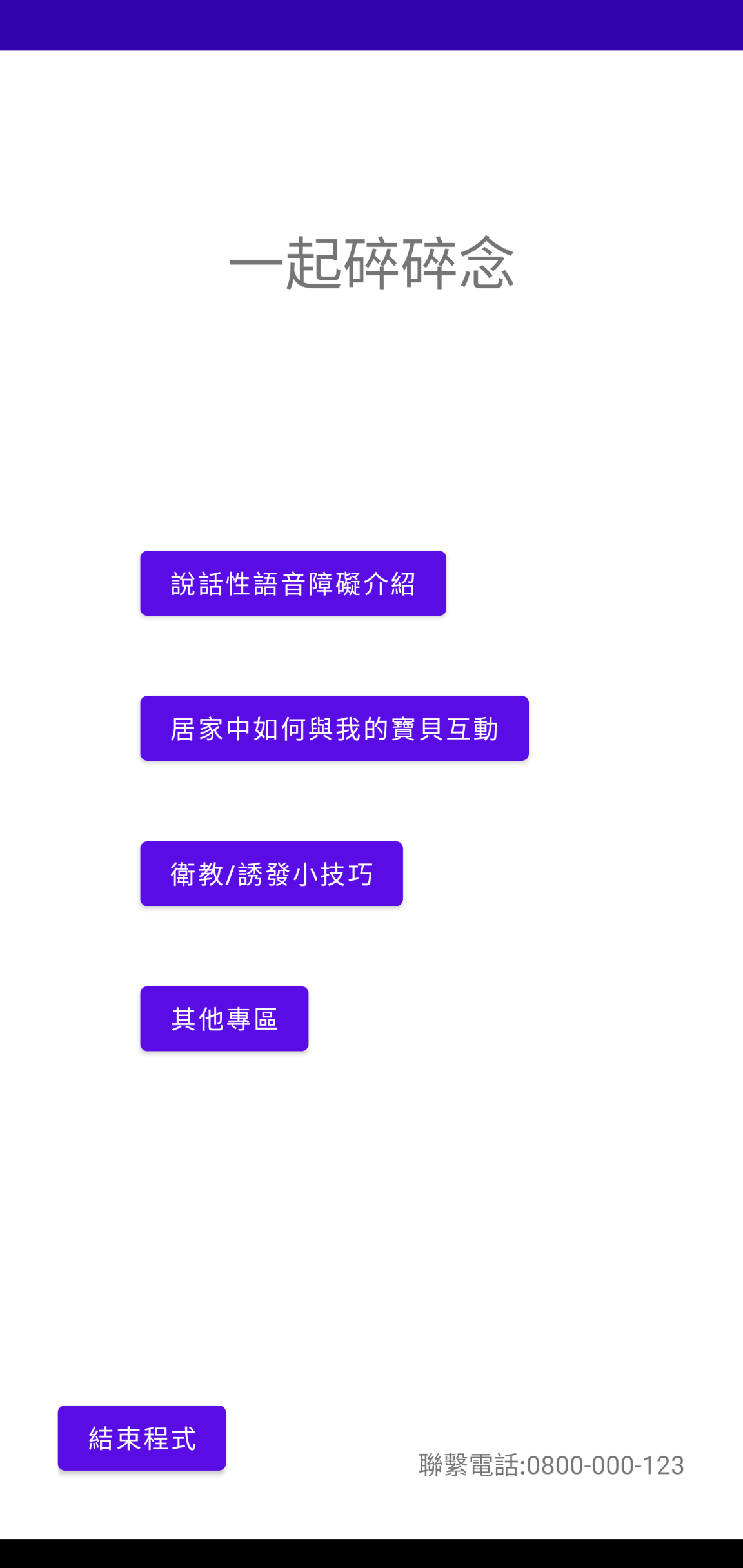} \\
			\small (\textbf{f})                     
			\label{fig:6}                           
		\end{tabular}
		\label{fig:images}
	\end{adjustwidth}
	\caption{Screenshot of the mobile~application: (\textbf{a}) Application main page: Users have the option of initiating a test or downloading the most recent classification model.; (\textbf{b}) User questionnaire page: The first step in the test procedure is to complete the questionnaire, which is primarily used for SSD background checks, such as whether the vocal organs are normal, etc.; (\textbf{c}) Recording interface: The recording session is the following step. The user can record by clicking the microphone button or play the sound file by clicking the play button.; (\textbf{d}) Analysis page: There are pages awaiting analysis.; (\textbf{e}) Result report page: Results page gives results for different SSDs categories.; (\textbf{f}) Health information link page: Links for Health Education and Promotion.\label{Screenshot}}
\end{figure}

\begin{adjustwidth}{-\extralength}{0cm}

	\reftitle{References}

\end{adjustwidth}

\begin{thebibliography}{999}
		
		\bibitem[Black \em{et~al.}(2015)Black, Vahratian, and
		Hoffman]{black2015communication}
		Black, L.I.; Vahratian, A.; Hoffman, H.J.
		\newblock {\em Communication Disorders and Use of Intervention Services among Children Aged 3--17 Years: United States, 2012};  Atlanta, GA, USA: US Department of Health and Human Services, Centers for Disease Control and Prevention, National Center for Health Statistics;
		\newblock 2015. 
		
		\bibitem[Wren \em{et~al.}(2016)Wren, Miller, Peters, Emond, and
		Roulstone]{wren2016prevalence}
		Wren, Y.; Miller, L.L.; Peters, T.J.; Emond, A.; Roulstone, S.
		\newblock Prevalence and predictors of persistent speech sound disorder at
		eight years old: Findings from a population cohort study.
		\newblock {\em J. Speech Lang. Hear. Res.} {\bf 2016},
		{\em 59},~647--673.
		
		\bibitem[The American Speech-Language-Hearing Association(2022)]{ASHA}
		Speech Sound Disorders-Articulation and Phonology, The American Speech-Language-Hearing Association. Accessed on 1 Jan 2022 from www.asha.org/practice-portal/clinical-topics/articulation-and-phonology.
		
		\bibitem[Chang and Yeh(2019)]{Chang2019Assessment}
		Chang, Y.N.; Yeh, L.L.
		\newblock Assessment Practices Followed by Speech-Language Pathologists for
		Clients with Suspected Speech Sound Disorder in Taiwan: A Survey Study.
		\newblock {\em Taiwan J. Phys. Med. Rehabil.} {\bf
			2019}, {\em 47},~31--47.
		
		\bibitem[Sen and Wang(2017)]{Sen2017Study}
		Sen, P.H.; Wang, C.L.
		\newblock {\em A Study of the Supply and Demand of Speech-Language Pathologist Manpower in Taiwan};
		\newblock Taiwan; University of Taipei; 2017. 
		
		\bibitem[Rvachew(2007)]{rvachew2007phonological}
		Rvachew, S.
		\newblock {\em Phonological Processing and Reading in Children with Speech Sound Disorders}; American Journal of Speech-Language Pathology; 2007. 
		
		\bibitem[Eadie \em{et~al.}(2015)Eadie, Morgan, Ukoumunne, Ttofari~Eecen, Wake,
		and Reilly]{eadie2015speech}
		Eadie, P.; Morgan, A.; Ukoumunne, O.C.; Ttofari~Eecen, K.; Wake, M.; Reilly, S.
		\newblock Speech sound disorder at 4 years: Prevalence, comorbidities, and
		predictors in a community cohort of children.
		\newblock {\em Dev. Med. Child Neurol.} {\bf 2015}, {\em
		57},~578--584.
		
		\bibitem[Jeng(2011)]{jeng2011phonological}
		Jeng, J.
		\newblock The phonological processes of syllable-initial consonants spoken by
		the preschool children of Mandarin Chinese.
		\newblock {\em J. Spec. Educ.} {\bf 2011}, {\em 34},~135--169.
		
		\bibitem[Anjos \em{et~al.}(2019)Anjos, Marques, Grilo, Guimar{\~a}es,
		Magalh{\~a}es, and Cavaco]{anjos2019sibilant}
		Anjos, I.; Marques, N.; Grilo, M.; Guimar{\~a}es, I.; Magalh{\~a}es, J.;
		Cavaco, S.
		\newblock Sibilant consonants classification with deep neural networks.
		\newblock In Proceedings of the EPIA Conference on Artificial Intelligence, Vila Real, Portugal, 3–6 September 2019; pp. 435--447.
		
		\bibitem[Krecichwost \em{et~al.}(2021)Krecichwost, Mocko, and
		Badura]{krecichwost2021automated}
		Krecichwost, M.; Mocko, N.; Badura, P.
		\newblock Automated detection of sigmatism using deep learning applied to
		multichannel speech signal.
		\newblock {\em Biomed. Signal Process. Control} {\bf 2021}, {\em
		68},~102612.
		
		\bibitem[Hammami \em{et~al.}(2020)Hammami, Lawal, Bedda, and
		Farah]{hammami2020recognition}
		Hammami, N.; Lawal, I.A.; Bedda, M.; Farah, N.
		\newblock Recognition of Arabic speech sound error in children.
		\newblock {\em Int. J. Speech Technol.} {\bf 2020}, {\em
		23},~705--711.
		
		\bibitem[Wang \em{et~al.}(2018)Wang, Chen, Yang, Dong, Xu, and
		Xu]{wang2018semi}
		Wang, F.; Chen, W.; Yang, Z.; Dong, Q.; Xu, S.; Xu, B.
		\newblock Semi-supervised disfluency detection.
		\newblock In Proceedings of the Proceedings of the 27th International
		Conference on Computational Linguistics, Santa Fe, NM, USA, 20--26 August 2018; pp. 3529--3538.
		
		\bibitem[Lou \em{et~al.}(2018)Lou, Anderson, and Johnson]{lou2018disfluency}
		Lou, P.J.; Anderson, P.; Johnson, M.
		\newblock Disfluency detection using auto-correlational neural networks.
		\newblock {\em arXiv} {\bf 2018}, arXiv:1808.09092.
		
		\bibitem[Wang \em{et~al.}(2017)Wang, Che, Zhang, Zhang, and
		Liu]{wang2017transition}
		Wang, S.; Che, W.; Zhang, Y.; Zhang, M.; Liu, T.
		\newblock Transition-based disfluency detection using lstms.
		\newblock In Proceedings of the 2017 Conference on Empirical
		Methods in Natural Language Processing, Copenhagen, Denmark, 7–11 September 2017; pp. 2785--2794.
		
		\bibitem[Tan and Le(2019)]{tan2019efficientnet}
		Tan, M.; Le, Q.
		\newblock Efficientnet: Rethinking model scaling for convolutional neural
		networks.
		\newblock In Proceedings of the International Conference on Machine Learning.
		PMLR,  Long Beach, CA, USA, 9--15 June 2019; pp. 6105--6114.
		
		\bibitem[Iandola \em{et~al.}(2014)Iandola, Moskewicz, Karayev, Girshick,
		Darrell, and Keutzer]{iandola2014densenet}
		Iandola, F.; Moskewicz, M.; Karayev, S.; Girshick, R.; Darrell, T.; Keutzer, K.
		\newblock Densenet: Implementing efficient convnet descriptor pyramids.
		\newblock {\em arXiv} {\bf 2014}, arXiv:1404.1869.
		
		\bibitem[Xia \em{et~al.}(2017)Xia, Xu, and Nan]{xia2017inception}
		Xia, X.; Xu, C.; Nan, B.
		\newblock Inception-v3 for flower classification.
		\newblock In Proceedings of the 2017 2nd International Conference on Image,
		Vision and Computing (ICIVC), Chengdu, China, 2--4 June 2017; pp. 783--787.
		
		\bibitem[Palanisamy \em{et~al.}(2020)Palanisamy, Singhania, and
		Yao]{palanisamy2020rethinking}
		Palanisamy, K.; Singhania, D.; Yao, A.
		\newblock Rethinking cnn models for audio classification.
		\newblock {\em arXiv} {\bf 2020}, arXiv:2007.11154.
		
		\bibitem[Nanni \em{et~al.}(2020)Nanni, Maguolo, and Paci]{nanni2020data}
		Nanni, L.; Maguolo, G.; Paci, M.
		\newblock Data augmentation approaches for improving animal audio
		classification.
		\newblock {\em Ecol. Inform.} {\bf 2020}, {\em 57},~101084.
		
		\bibitem[Tomar(2006)]{tomar2006converting}
		Tomar, S.
		\newblock Converting video formats with FFmpeg.
		\newblock {\em Linux J.} {\bf 2006}, {\em 2006},~10.
		
		\bibitem[McFee \em{et~al.}(2015)McFee, Raffel, Liang, Ellis, McVicar,
		Battenberg, and Nieto]{mcfee2015librosa}
		McFee, B.; Raffel, C.; Liang, D.; Ellis, D.P.; McVicar, M.; Battenberg, E.;
		Nieto, O.
		\newblock librosa: Audio and music signal analysis in python.
		\newblock In Proceedings of the 14th Python in Science
		Conference, Austin, TX, USA, 6--12 July 2015; Volume~8, pp. 18--25.
		
		\bibitem[Abadi \em{et~al.}(2015)Abadi, Agarwal, Barham, Brevdo, Chen, Citro,
			Corrado, Davis, Dean, Devin, Ghemawat, Goodfellow, Harp, Irving, Isard,
			Jozefowicz, Jia, Kaiser, Kudlur, Levenberg, Mané, Schuster, Monga, Moore,
			Murray, Olah, Shlens, Steiner, Sutskever, Talwar, Tucker, Vanhoucke,
			Vasudevan, Viégas, Vinyals, Warden, Wattenberg, Wicke, Yu, and
		Zheng]{Abadi_TensorFlow_Large-scale_machine_2015}
		Abadi, M.; Agarwal, A.; Barham, P.; Brevdo, E.; Chen, Z.; Citro, C.; Corrado,
		G.S.; Davis, A.; Dean, J.; Devin, M.;  et~al.
		\newblock {TensorFlow, Large-Scale Machine Learning on Heterogeneous Systems}. www.tensorflow.org/, Accessed on 1 Jan 2022,
		2015.  https://doi.org/10.5281/zenodo.4724125.
		
	\end{thebibliography}
\end{document}